\documentclass[11pt]{iopart}
\usepackage{iopams}
\usepackage{amssymb}
\usepackage{graphicx}
\usepackage{color}
\usepackage[tight,TABTOPCAP]{subfigure}
\usepackage{braket}
\usepackage{epstopdf}
\begin{document}
\makeatletter

\newbox\slashbox \setbox\slashbox=\hbox{$/$}
\newbox\Slashbox \setbox\Slashbox=\hbox{$/$}
\def\pFMslash#1{\setbox\@tempboxa=\hbox{$#1$}
  \@tempdima=0.5\wd\slashbox \advance\@tempdima 0.5\wd\@tempboxa
  \copy\slashbox \kern-\@tempdima \box\@tempboxa}
\def\pFMSlash#1{\setbox\@tempboxa=\hbox{$#1$}
  \@tempdima=0.5\wd\Slashbox \advance\@tempdima 0.5\wd\@tempboxa
  \copy\Slashbox \kern-\@tempdima \box\@tempboxa}
\def\FMslash{\protect\pFMslash}
\def\FMSlash{\protect\pFMSlash}
\def\miss#1{\ifmmode{/\mkern-11mu #1}\else{${/\mkern-11mu #1}$}\fi}
\makeatother


\title[Diphoton Higgs signal strength...]{Diphoton Higgs signal strength in universal extra dimensions}
\author{I. Garc\'\i a-Jim\' enez$^{(a)}$,  J. Monta\~no$^{(b),(c)}$, G. N\' apoles-Ca\~nedo$^{(d)}$, H. Novales-S\' anchez$^{(d)}$,  J. J. Toscano$^{(d)}$, and E. S. Tututi$^{(c)}$}
\address{$^{(a)}$Departamento de Ciencias B\' asicas, Instituto Tecn\' ologico de Oaxaca, Av. Ing. Victor Bravo Ahuja No. 125 Esquina Calzada Tecnol\' ogico C. P. 68030, Oaxaca, Oaxaca, M\' exico. \\
$^{(b)}$CONACYT, Av. Insurgentes Sur 1582, Col. Cr\'edito Constructor, Alc. Benito Ju\'arez,
C.~P. 03940, Ciudad de M\'exico, M\'exico.\\
$^{(c)}$Facultad de Ciencias F\'\i sico Matem\' aticas,
Universidad Michoacana de San Nicol\'as de
Hidalgo, Avenida Francisco J. M\'ujica S/N, 58060, Morelia, Michoac\'an, M\' exico. \\
$^{(d)}$Facultad de Ciencias F\'{\i}sico Matem\'aticas,
Benem\'erita Universidad Aut\'onoma de Puebla, Apartado Postal
1152, Puebla, Puebla, M\'exico.}

\begin{abstract}
The signal strength of the $gg \to H \to \gamma \gamma$ reaction in $pp$ collisions at the Large Hadron Collider is studied within the context
of the Standard Model with universal extra dimensions (UED). The impact of an arbitrary number $n$ of UED on both the $gg\to H$ and $H\to \gamma \gamma$ subprocesses is studied. The one-loop contribution of Kaluza-Klein excitations on these subprocesses are proportional to discrete and continuous sums, $\sum_{(\underline{k})}\int d^4k$, which can diverge. By implementing the dimensional regularization scheme, it is shown that discrete regularized sums can naturally be expressed as multidimensional Epstein functions, and that divergences, if exist, emerge through the poles of these functions. It is found that continuous sums converge, but the discrete ones diverge, with the exception of the $n=1$ case, in which the one-dimensional Epstein function converges. It is argued that divergences that arise from discrete sums for $n\geq 2$ are genuine ultraviolet divergences, since they correspond to short-distance effects in the compact manifold. Then, the amplitudes are renormalized in a modern sense by incorporating interactions of canonical dimension higher than four that allow us to generate the required counterterms, which are determined using a $\overline{\rm MS}$-like renormalization scheme. We find that the $gg\to H$ subprocess is quite sensitive to both the size and the dimension of the compact manifold, but the Standard Model prediction for $H\to \gamma \gamma$ subprocess is practically unchanged. In the $n=1$ case, it is found that the experimental constraint on the compactification scale $R^{-1}\geq 1.5\, TeV$ allow us to reproduce the experimental limit on the signal strength $1.01\leq \mu^{(1)}_{\gamma \gamma}\leq 1.2$. In the $n\geq 2$ cases, it is found that the experimental limit on $\mu^{(n)}_{\gamma \gamma}$ leads to stronger lower bounds for the compactification scale given by $R^{-1}\geq 1.55, \, 2.45, \, 3.57, \, 5.10, \, 7.25$ TeVs for $n=2,\, 4,\, 6,\, 7,\, 8,\, 10$, respectively.

\end{abstract}


\maketitle

\section{Introduction}
\label{In}The Higgs mechanism~\cite{HM1,HM2,HM3,HM4,HM5,HM6} is a key building block of the Standard Model (SM). The Higgs boson is the particle predicted to exist as a consequence of implementing the Higgs mechanism in the electroweak sector of the model. This particle plays a unique role in the SM, as it explains the origin of mass. The discovery in 2012 of a scalar resonance by the ATLAS~\cite{ATLAS} and CMS~\cite{CMS} Collaborations at the Large Hadron Collider (LHC)~\cite{LHC}, with features that resemble the SM Higgs boson, opened a new era in high-energy Higgs physics. In order to establish unambiguously that this scalar resonance actually corresponds to the Higgs boson predicted by the SM, many of the forthcoming experiments in LHC, as well as those planed to be performed in the International Linear Collider~\cite{ILC}, will be focused in studying its decays into SM particles. In particular, a great deal of experimental effort will be dedicated to measure, with a high level of precision, the Higgs boson decay into two photons, as this process is very sensitive to new-physics effects. Since this decay is naturally suppressed in the sense that it first arises at the one-loop level in any renormalizable theory, it constitutes a promising process to look for sources of new physics in general. The ATLAS and CMS Collaborations have reported signal strengths in diverse channels~\cite{PDG}. In particular, the signal strength in the decay channel $\gamma \gamma$ relative to the SM prediction is~\cite{PDG,ATLAS1,CMS1,Tevatron}:
\begin{equation}
\label{cs1}
\mu_{\gamma \gamma }=\frac{\sigma(pp\to H \to \gamma \gamma)}{\sigma(pp\to H \to \gamma \gamma)^{\rm SM}}=1.10^{+0.10}_{-0.09}\, .
\end{equation}
A precise measurement of $\mu_{\gamma \gamma}$ is of great importance because any deviation of it from unity would suggest the presence of new-physics (NP) effects.  Assuming that gluon fusion is the dominant channel of Higgs production and using the narrow width approximation, the above expression becomes
\begin{equation}
\mu_{\gamma \gamma }=\frac{\sigma(gg\to H)\times {\rm BR}(H\to \gamma \gamma)}{\sigma(gg\to H)^{\rm SM}\times {\rm BR}(H\to \gamma \gamma)^{\rm SM}}\, .
\end{equation}
The cross section $\sigma(gg\to H)$ has a strong correlation with the decay width $\Gamma(H\to gg)$, so that the above expression can be written as follows:
\begin{equation}
\label{csw}
\mu_{\gamma \gamma }=\frac{\Gamma(H\to gg)\times {\rm BR}(H\to \gamma \gamma)}{\Gamma(H\to gg)^{\rm SM}\times {\rm BR}(H\to \gamma \gamma)^{\rm SM}}\, .
\end{equation}
Assuming that the production cross section is that of the SM Higgs boson, that is, $\Gamma(H\to gg)=\Gamma(H\to gg)^{\rm SM}$, constraints can be derived on the quantity ${\rm BR}(H\to \gamma \gamma)$ in any theoretical model from
\begin{equation}
\mu_{\gamma \gamma }=\frac{{\rm BR}(H\to \gamma \gamma)^{\rm NP}}{{\rm BR}(H\to \gamma \gamma)^{\rm SM}}\, .
\end{equation}
The behavior of this relation has been investigated by several authors in various models beyond the SM, such as the Inert Higgs Doublet Model~\cite{IHDM}, models with two and three Higgs Doublets~\cite{THDM}, 331 models~\cite{331M}, Higgs triplet models~\cite{HTM}, more general extended Higgs sectors~\cite{EHS}, and in a model-independent manner, using the effective-Lagrangian technique~\cite{EL1,EL2}.\\

More important than the branching ratios, however, is the total cross section of $\sigma(pp\to H\to \gamma \gamma)$, since that is what is measured at the collider. In this paper, we are interested in studying the signal strength given by Eq.~(\ref{csw}) in the context of compact extra dimensions~\cite{A, AHA}. We will focus on the so-called Universal Extra Dimensions (UED) models~\cite{ACD}. We will present our study within the context of a SM extension that incorporates $n$ UED~\cite{EDSM}. From now on, we will refer to this extra-dimensional version of the SM with the acronym EDSM. We will assume that the number of spatial extra dimensions is even\footnote{The only exception will be the case of one extra dimension, as it presents some interesting features that deserve special attention.}, that is $n=2, 4, \cdots$, since chirality is well defined in spacetimes of even dimension, which in turn facilitates the recovery of the chiral SM structure after compactification. In this regard, the $H\to \gamma \gamma$ decay in models with extra dimensions has already been the subject of important attention. It was first studied by Petriello~\cite{Petriello} in the context of the SM with one UED. Diverse studies of this decay have been carried out before~\cite{BHD} and after~\cite{AHD,AHD6D} the Higgs boson discovery by considering diverse geometries of the compact manifold.\\

Our main goal in this work is to investigate the behavior of the $\mu_{\gamma \gamma}$ signal strength for arbitrary even values of the number of extra dimensions $n$. One important feature of models of UED is that contributions of Kaluza-Klein (KK) excitations to standard observables first arise at the one-loop level. This fact is a consequence of the $(4+n)$-dimensional momentum conservation, which is known in the literature as KK parity conservation. At energies much higher than the compactification scale $R^{-1}$, with $R$ the size of the compact manifold, one starts from an effective field theory governed by the extended Poincar\' e group, ${\rm ISO}(1,3+n)$, and the extended SM gauge group, $G({\cal M}^{4+n})\equiv {\rm SU}_C(3,{\cal M}^{4+n})\times {\rm SU}_L(2,{\cal M}^{4+n})\times {\rm U}_Y(1,{\cal M}^{4+n})$, where ${\cal M}^{4+n}={\cal M}^4\times {\cal N}^n$, with ${\cal M}^4$ the usual four-dimensional manifold and ${\cal N}^n$ a flat $n$-dimensional Euclidean manifold. The most general effective Lagrangian that respects this symmetry can be written as: ${\cal L}_{\rm eff}(x,\bar x)={\cal L}^{\rm SM}_{(\mathbf{d}\leq 4+n)}(x,\bar x)+{\cal L}_{(\mathbf{d}> 4+n)}(x,\bar x)$, with $x\in {\cal M}^4$ and $\bar x \in {\cal N}^4$. In this Lagrangian, the first term corresponds to the $(4+n)$-dimensional version of the SM, which contains interactions of canonical dimension less than or equal to $(4+n)$, while the second term includes all interactions of canonical dimension higher than $(4+n)$ that are compatible with these symmetries, since the theory is nonrenormalizable in the usual sense. From dimensional considerations, the ${\rm ISO}(1,3+n)\times G({\cal M}^{4+n})$-invariant interactions that appear in the ${\cal L}_{(\mathbf{d}> 4+n)}(x,\bar x)$ Lagrangian will be multiplied  by inverse powers of a very high energy scale $\Lambda \gg R^{-1}$, so they will be naturally suppressed with respect to the SM-like interactions that constitute the ${\cal L}^{\rm SM}_{(\mathbf{d}\leq 4+n)}(x,\bar x)$ Lagrangian, which do not depend on this scale. This means that this Lagrangian determines the dominant contributions to physical observables.\\

At energies below the compactification scale, we need to go from the ${\rm ISO}(1,3+n)\times G({\cal M}^{4+n})$ description to a description based on the usual ${\rm ISO}(1,3)\times G({\cal M}^4)$ groups, for which it is necessary to implement a procedure of hiding of symmetry. Hiding the ${\rm ISO}(1,3+n)\times G({\cal M}^{4+n})$ symmetry into the ${\rm ISO}(1,3)\times G({\cal M}^{4})$ symmetry requires of implementing two canonical maps, one that accommodates covariant objects of ${\rm SO}(1,3+n)$ into covariant objets of ${\rm SO}(1,3)$, the other allowing us to hide, through a general Fourier series, any dynamical role of the ${\rm ISO}(n)$ subgroup of ${\rm ISO}(3+n)$~\cite{OP1,OP2,OP3,EDQED,EDYM}. Note that the $ G({\cal M}^{4+n})$ and  $G({\cal M}^{4})$ groups coincide as Lie groups since they have the same number of generators, but they differ as gauge groups because their support manifolds are different. As already commented, the  ${\cal L}^{\rm SM}_{(\mathbf{d}\leq 4+n)}(x,\bar x)$ Lagrangian plays, from a phenomenological point of view, a dominant role at low energies. In fact, after compactification and integration on the $\bar x$ coordinates, the ${\cal L}^{\rm SM}_{(\mathbf{d}\leq 4+n)}(x,\bar x)$ Lagrangian unfolds into two terms, one that corresponds to the usual four-dimensional theory, ${\cal L}^{(\underline{0})}_{\rm SM}$, and the other that contains interactions between usual fields (zero-mode fields) and KK excitations, ${\cal L}^{\rm KK}_{(\mathbf{d}\leq 4)}$. The important point to be emphasized is that, like ${\cal L}^{(\underline{0})}_{\rm SM}$, the ${\cal L}^{\rm KK}_{(\mathbf{d}\leq 4)}$ Lagrangian contains only interactions of canonical dimension less than or equal to four, so they cannot depend on the high energy scale $\Lambda$. This means that the ${\cal L}^{\rm KK}_{(\mathbf{d}\leq 4)}$ Lagrangian only can depend on the compactification scale $R^{-1}$ through masses of the KK excitations. Of course, physical amplitudes will also depend on the number $n$ of extra dimensions. This is the reason why this Lagrangian has a dominant role in phenomenological predictions, even though its effects first appear at the one-loop level. Therefore, if we are interested in making phenomenological predictions, it is clear that we will face both conceptual and technical complications that go beyond the conventional formalism of radiative corrections.\\

 Our objective is twofold. On the one hand, we are interested in investigating the sensitivity of this signal strength to the size and to the dimension of the compact manifold. On the other hand, we are interested in using this process to study some technical aspects concerning radiative corrections in the context of this class of effective theories, which are intrinsically nonrenormalizable in the Dyson's sense. As it is discussed in previous communications by some of us~\cite{EDQED,EDYM,EDST}, the main challenge we face in obtaining loop predictions of Kaluza-Klein (KK) theories is not that they are effective theories nonrenormalizable in the usual sense, but the fact that they have an infinite number of fields. As a consequence, the KK excited modes give one-loop contributions that are proportional to discrete and continuous infinite sums, $\sum_{(\underline{k})}\int d^4k$, with $\sum_{(\underline{k})}$ a short-hand notation for a total of $2^n-1$ nested infinite sums (see the next section for notation and conventions). Both discrete and continuous sums can diverge, so they need to be regularized. Through the use of the dimensional regularization scheme~\cite{BoGi,TVreg}, it has been shown in~\cite{EDQED,EDYM,EDST} that both continuous and discrete sums can simultaneously be regularized, and that the latter can naturally be expressed in terms of multidimensional Epstein functions~\cite{E1}. So divergences that arise from continuous and discrete sums appear as poles of the gamma and Epstein functions, respectively. As it is argued in~\cite{EDQED,EDYM,EDST}, divergences arising from discrete sums are genuine ultraviolet divergences, since they are linked to short-distance effects in the compact manifold, and thus they can be removed from physical amplitudes by renormalization in a modern sense~\cite{Weinberg}. So, in this type of theories there are two types  of ultraviolet divergences because there are two different spaces, namely the usual infinite manifold ${\cal M}^4$ and the compact manifold. If only one-loop effects of the ${\cal L}^{(\underline{0})}_{\rm SM}$ and ${\cal L}^{\rm KK}_{(\mathbf{d}\leq 4)}$ Lagrangians are considered, two scenarios can arise when studying usual Green's functions (Green's functions whose external legs correspond all to zero modes):\\

 \noindent $\bullet$ \textit{Interactions of canonical dimension less than or equal to 4}. In this case, the one-loop contributions of the KK excited-modes to the usual fields and parameters (constant couplings and masses) are proportional to products of the way $\Gamma\left(\frac{\epsilon}{2}\right)E^{c^2}_l\left(\frac{\epsilon}{2}\right)$, with $\Gamma\left(\frac{\epsilon}{2}\right)$ the gamma function, $E^{c^2}_l\left(\frac{\epsilon}{2}\right)$ the $l$-dimensional Epstein function, and $\epsilon=4-D$ the complex parameter introduced through the dimensional regularization scheme. Since the $l$-dimensional Epstein function $E^{c^2}_l(s)$ has poles at $s=\frac{l}{2}, \frac{l-1}{2}, \cdots, -\frac{1}{2},-\frac{3}{2},-\cdots $, except for zero~\cite{K}, it is clear that $E^{c^2}_l\left(\frac{\epsilon}{2}\right)$ converges when $\epsilon$ goes to zero. However, we must be careful when taking this limit in the $\Gamma\left(\frac{\epsilon}{2}\right)E^{c^2}_l\left(\frac{\epsilon}{2}\right)$ product, since the gamma function diverges in such a limit. As it is shown in Refs.~\cite{EDQED,EDYM}, the decomposition of the $l$-dimensional Epstein function in terms of the one-dimensional Epstein function~\cite{E3}, allows us to identify two types of ultraviolet divergences, one associated with short-distance effects in the usual spacetime manifold, the other linked to short-distance effects in the compact manifold. To see this note that
 \begin{eqnarray}
 \label{C1}
&&\Gamma\left(\frac{\epsilon}{2}\right) E^{c^2}_l\left(\frac{\epsilon}{2}\right)=\Gamma\left(\frac{\epsilon}{2}\right) \Bigg[E^{c^2}_1\left(\frac{\epsilon}{2}\right)\nonumber \\
&& + \frac{(-1)^{l-1}}{2^{l-1}}\sum^{l-1}_{p=1}\left(\begin{array}{ccc}
l-1 \\
p
\end{array}\right)(-1)^p \pi^{\frac{p}{2}}\frac{\Gamma\left(\frac{\epsilon}{2}-\frac{p}{2}\right)}
{\Gamma\left(\frac{\epsilon}{2}\right)}E^{c^2}_1\left(\frac{\epsilon}{2}-\frac{p}{2}\right)\Bigg]\, .
\end{eqnarray}
Two types of divergences can be identified from this expression. Since the one-dimensional Epstein function $E^{c^2}_1\left(\frac{\epsilon}{2}\right)$ converges when $\epsilon \to 0$,  the first term of the above expression diverges exclusively through the pole of the gamma function in this limit. Since the gamma function emerges from the continuous sum, that is, $\int d^Dk \sim \Gamma\left(\frac{\epsilon}{2}\right)$, this type of divergences must correspond to usual divergences in the sense that they arise from short-distance effects of the KK excited-modes in the spacetime manifold ${\cal M}^4$. The $E^{c^2}_1\left(\frac{\epsilon}{2}\right)$ factor, which does not depend on $c^2$ but only on the number of extra dimensions, quantify, in the $\epsilon \to 0$ limit, the ultraviolet contribution of the infinite number of KK excitations. Another type of divergences arises from the sum over the $p$ index. In this case, we can see that the $\Gamma\left(\frac{\epsilon}{2}\right)$ function is canceled by the denominator appearing in this sum, so divergences emerge either as poles of the gamma function or of the Epstein function, which appear in this sum as products of the form
$\Gamma\left(\frac{\epsilon}{2}-\frac{p}{2}\right) E^{c^2}_1\left(\frac{\epsilon}{2}-\frac{p}{2}\right)$. This type of divergences, which depend polynomially on the external momentum normalized to the compactification  scale through the $c^2$ function, arise as poles of the gamma function for even values of $p$ or as poles of the Epstein function for odd values of $p$, since the one-dimensional Epstein function $E^{c^2}_1(s)$ has poles at $s=\frac{1}{2},-\frac{1}{2},-\frac{3}{2}, \cdots$, except for zero~\cite{K}. This is the type of ultraviolet divergences which we have identified as short-distance effects in the compact manifold~\cite{EDQED,EDYM}, since they arise from the Epstein function or, equivalently, from the discrete infinite sum. The link between the compact manifold and the discrete infinite sum is given via Fourier series, in the same way that the infinite manifold and the continuous sum are connected through Fourier transform. While, the former is a sum on discrete momenta, the latter is a sum on continuous momenta. However, note that this type of divergences does not exist in the five-dimensional theory, since in this case $l=1$. Since these divergences arise as coefficients of polynomial expressions in external momentum, appropriate interactions of canonical dimension higher then four, which already are present in the effective theory, must be considered in order to generate the needed counterterms to remove them from physical amplitudes~\cite{EDQED,EDYM}.\\

\noindent $\bullet$ \textit{Interactions of canonical dimension greater than four}. In this case, the one-loop amplitudes induced by the ${\cal L}^{(\underline{0})}_{\rm SM}$ and ${\cal L}^{\rm KK}_{(\mathbf{d}\leq 4)}$ Lagrangians are proportional to terms of the way $\Gamma\left(m+\frac{\epsilon}{2}\right)E^{c^2}_l\left(m+\frac{\epsilon}{2}\right)$, with $m$ a positive integer ($m\geq 1$). It is clear that the gamma function  $\Gamma\left(m+\frac{\epsilon}{2}\right)$ always converges in this case, so divergences, if exist, will arise from poles of the $l$-dimensional Epstein function. This means that any divergence should be associated with a short-distance effect in the compact manifold, since it arises from discrete infinite sums. In other words, no short-distance effects in the usual spacetime manifold can arise, since the usual continuous sum converges in this case. To see this, we express the multidimensional Epstein function in terms of the one-dimensional Epstein function to obtain an analogous expression of (\ref{C1}) given by
\begin{eqnarray}
 \label{C2}
&&\Gamma\left(m+\frac{\epsilon}{2}\right) E^{c^2}_l\left(m+\frac{\epsilon}{2}\right)=\Gamma\left(m+\frac{\epsilon}{2}\right) \Bigg[E^{c^2}_1\left(m+\frac{\epsilon}{2}\right)\nonumber \\
&& + \frac{(-1)^{l-1}}{2^{l-1}}\sum^{l-1}_{p=1}\left(\begin{array}{ccc}
l-1 \\
p
\end{array}\right)(-1)^p \pi^{\frac{p}{2}}\frac{\Gamma\left(m+\frac{\epsilon}{2}-\frac{p}{2}\right)}
{\Gamma\left(m+\frac{\epsilon}{2}\right)}E^{c^2}_1\left(m+\frac{\epsilon}{2}-\frac{p}{2}\right)\Bigg].\, \nonumber \\
&&
\end{eqnarray}
As anticipated, the first terms of this expression tell us that there are no short-distance effects in the usual spacetime manifold, since both $\Gamma\left(m+\frac{\epsilon}{2}\right)$ and $E^{c^2}_1\left(m+\frac{\epsilon}{2}\right)$ converge as $\epsilon \to 0$. Divergences arise from terms summed over the $p$ index as poles of the gamma function and of the one-dimensional Epstein function for even and odd values of $p$, respectively. This type of ultraviolet divergences are associated with short-distance effects in the compact manifold, so interactions of canonical dimension higher than four must be considered in order to generate the needed counterterms to define renormalized amplitudes. However, note that in the special case of only one extra dimension, this type of divergences is absent, so any interaction of canonical dimension higher than four is free of ultraviolet divergences, a fact well known in the literature~\cite{ACD,Petriello,P5D}. In the case at hand, the $H\gamma \gamma$ coupling has canonical dimension 5, so $m=1$. This coupling is free of ultraviolet divergences for the special case $n=1$~\cite{Petriello}, but it diverges for $n\geq 2$. For comparison purposes, we will present results for the $H\to gg$ and $H\to \gamma \gamma$ decays in both the $n=1$ and $n\geq 2$ cases. We will put special attention on how divergences can be regularized and removed from physical amplitudes, as we think that a consistent theoretical scheme to calculate electroweak observables in this context is of great interest from both the phenomenological and the experimental points of view.

The rest of the paper has been organized as follows. In Sec.~\ref{Mo}, we follow Refs.~\cite{EDSM,EDQED,EDYM,OP1,OP2,OP3,OP4,PVSM} to present a quite schematic description of the EDSM, which includes the Feynman rules that are needed for our calculations. Sec.~\ref{Ca} is devoted to calculate the widths of the $H\to \gamma \gamma$ and $H\to gg$ decays in the context of the EDSM. Several scenarios for an even number of extra dimensions are studied and compared with the case $n=1$. Finally, in Sec.~\ref{Co} the conclusions are presented.

\section{Model outline}
\label{Mo}In this section, we present a brief discussion on the SM with universal extra dimensions. Our discussion will be based on Refs.~\cite{EDSM,EDQED,EDYM,OP1,OP2,OP3,OP4,PVSM}, avoiding, as much as possible, technical details, which can be found in these references.

\subsection{Hiding the symmetry}
As already commented in the introduction, our starting point at very high energies is an effective field theory that is governed by the extended ${\rm ISO}(1,3+n)\times G({\cal M}^{4+n})$ groups, which contains, besides the $(4+n)$-dimensional version of the SM, all interactions of canonical dimension higher than $4+n$ that respect this symmetry. At energies below the compactification scale, one hides the ${\rm ISO}(1,3+n)\times G({\cal M}^{4+n})$ symmetry into the usual ${\rm ISO}(1,3)\times G({\cal M}^{4})$ symmetry through two canonical maps. The idea behind these  maps is to hide the dynamic role of subgroup ${\rm ISO}(n)$ of ${\rm ISO}(1,3+n)$, since at low energies, order distances of the size of the extra coordinates are explored. One of these maps  accommodates covariant objets of the extended Lorentz group ${\rm SO}(1,3+n)$ into covariant objects of the standard Lorentz group ${\rm SO}(1,3)$. This means that a spinor $\Psi(x,\bar x)$ of ${\rm SO}(1,3+n)$ is mapped into $2^{\frac{n}{2}}$ spinors $\psi(x,\bar x)$ of ${\rm SO}(1,3)$, while a vector ${\cal A}_M(x,\bar x)$ of ${\rm SO}(1,3+n)$ is mapped into a 4-vector, ${\cal A}_\mu(x,\bar x)$, and $n$ scalar fields, ${\cal A}_{\bar \mu}(x,\bar x)$, of ${\rm SO}(1,3)$, where $M=\mu\, , \bar \mu$, with $\mu=0,1,2,3$ and $\bar \mu=5,\cdots, 4+n$. To completely hide the ${\rm ISO}(1,3+n)\times G({\cal M}^{4+n})$ symmetry into the ${\rm ISO}(1,3)\times G({\cal M}^{4})$ symmetry, a second canonical map is needed, since after the first map the diverse fields still depend on the extra coordinates. It should be remembered that, at very high energies, the $\bar x$ points on ${\cal N}^n$ label degrees of freedom, but this role must be hidden at low energies. At this energies, we need to choose a compactification procedure, but we can establish the essence of the low-energy theory without establishing a specific geometry. Assume that some compactification procedure on the ${\cal N}^n$ manifold has been carried out, and let $\{f^{(\underline{k})}(\bar x)\}$ be a complete set of orthogonal functions defined on the compact manifold. In the following, we will illustrate the main ideas using the gauge fields, ${\cal A}^a_M(x,\bar x) \mapsto \{{\cal A}^a_\mu(x,\bar x), {\cal A}^a_{\bar \mu}(x,\bar x)\}$, and the gauge parameters, $\alpha^a(x,\bar x)$, of a generic ${\rm SU}(N,{\cal M}^{4+n})$ gauge group. Then, these fields and parameters can be decomposed in this basis through a general Fourier series as follows:
\begin{eqnarray}
{\cal A}^a_\mu(x,\bar x)&=&\sum_{(\underline{k})}f^{(\underline{k})}(\bar x)A^{(\underline{k})a}_\mu(x)\, ,\\
{\cal A}^a_{\bar \mu}(x,\bar x)&=&\sum_{(\underline{k})}f^{(\underline{k})}(\bar x)A^{(\underline{k})a}_{\bar \mu}(x)\, ,\\
\alpha^a(x,\bar x)&=&\sum_{(\underline{k})}f^{(\underline{k})}(\bar x)\alpha^{(\underline{k})a}( x)\, ,
\end{eqnarray}
where the $A^{(\underline{k})}_\mu(x)$ and $A^{(\underline{k})}_{\bar \mu}(x)$ fields represent the degrees of freedom, while the $f^{(\underline{k})}(\bar x)$ functions do not represent degrees of freedom. The contact with the usual theory should not depend on the compactification procedure, which can be achieved by introducing the constant function $f^{(\underline{0})}$ into our basis. Then, we postulate that any field with standard counterpart must have a component along the $f^{(\underline{0})}$ direction, identifying such component as the zero mode field or usual field. This is the case of the gauge ${\cal A}^a_\mu(x,\bar x)$ fields and the gauge $\alpha^a(x,\bar x)$ parameters, so they must be expressed as follows:
\begin{eqnarray}
{\cal A}^a_\mu(x,\bar x)&=&f^{(\underline{0})}A^{(\underline{0}) a}_\mu(x)+\sum_{(\underline{k})}f^{(\underline{k})}(\bar x)A^{(\underline{k})a}_\mu(x)\, ,\\
\alpha^a(x,\bar x)&=&f^{(\underline{0})}\alpha^{(\underline{0})a}( x)+\sum_{(\underline{k})}f^{(\underline{k})}(\bar x)\alpha^{(\underline{k})a}( x)\, .
\end{eqnarray}
The compactification procedure does not generate mass terms for the $A^{(\underline{0}) a}_\mu(x)$ fields because such mechanism operates through derivatives, with respect to the $\bar x$ coordinates, of the basis functions. This allows us to identify the $A^{(\underline{0}) a}_\mu(x)$ fields and the $\alpha^{(\underline{0})a}( x)$ parameters as the gauge fields and gauge parameters of the standard ${\rm SU}(N,{\cal M}^4)$ gauge group. However, there are also some fields without standard counterpart, so they do not have component along the $f^{(\underline{0})}$ direction. This is the case of the scalar fields ${\cal A}^a_{\bar \mu}(x,\bar x)$. The fact that the $f^{(\underline{0})}$ function is trivially even under $\bar x \to -\bar x$ suggests that the basis $\{f^{(\underline{0})}, \, f^{(\underline{k})}(\bar x) \}$ can be reorganized into two independent basis, one containing the even functions, $\{f^{(\underline{0})}_E, \, f^{(\underline{k})}_E(\bar x) \}$, the other containing the odd functions, $\{ f^{(\underline{k})}_O(\bar x) \}$. So, the
scalar fields ${\cal A}^a_{\bar \mu}(x,\bar x)$ will be expressed as follows:
\begin{equation}
{\cal A}^a_{\bar \mu}(x,\bar x)=\sum_{(\underline{k})}f^{(\underline{k})}_O(\bar x)A^{(\underline{k})a}_{\bar \mu}(x)\, .
\end{equation}
 A defined parity of the basis functions is established by assuming that the compact manifold is made of $n$ replicas of the orbifold $S^1/Z_2$, where it is assumed, for simplicity, that the radii of the $n$ $S^1$ circles are all equal, that is, $R_1=\cdots=R_n\equiv R$. Then, one postulates that fields with standard counterpart are even under $\bar x\to -\bar x$, while those without standard counterpart are odd under this operation.\\

The original gauge transformation $\delta {\cal A}^a_M(x,\bar x)={\cal D}^{ab}_M \alpha^b(x,\bar x)\mapsto \{\delta {\cal A}^a_\mu(x,\bar x)={\cal D}^{ab}_\mu \alpha^b(x,\bar x), \delta {\cal A}^a_{\bar \mu}(x,\bar x)={\cal D}^{ab}_{\bar \mu} \alpha^b(x,\bar x)  \}$, leads to two types of gauge transformations, one defined by the zero-mode gauge parameters $\alpha^{(\underline{0})a}(x)$, which we call standard gauge transformations (SGTs), the other by the excited-mode gauge parameters $\alpha^{(\underline{k})a}(x)$, which, to distinguish them from the former, will be called nonstandard gauge transformations (NSGTs). The SGTs define the usual ${\rm SU}(N,{\cal M}^4)$ gauge group, under which the $A^{(\underline{0})a}_\mu(x)$ fields transform as gauge fields, whereas the $\{A^{(\underline{k})a}_\mu(x), A^{(\underline{k})a}_{\bar \mu}(x)\}$ fields, which are recognized as KK excitations of $A^{(\underline{0})a}_\mu(x)$, transform in the adjoint representation of this group. Since the $\{A^{(\underline{k})a}_\mu(x), A^{(\underline{k})a}_{\bar \mu}(x)\}$ fields appear as matter fields from the ${\rm SU}(N,{\cal M}^4)$ perspective, they can be endowed with mass. On the other hand, the NSGTs remind us that there is a larger gauge symmetry than the usual ${\rm SU}(N,{\cal M}^4)$ symmetry. This means that there are KK excitations of the $A^{(\underline{0})a}_\mu(x)$ gauge fields that are also gauge fields. However, not all the $\{A^{(\underline{k})a}_\mu(x), A^{(\underline{k})a}_{\bar \mu}(x)\}$ fields can be gauge fields, since there is one-to-one relation between gauge fields and gauge parameters. It is not difficult to convince ourselves that the only possibility is $A^{(\underline{k})a}_\mu(x) \leftrightarrow \alpha^{(\underline{k})a}(x)$, so the $A^{(\underline{k})a}_{\bar \mu}(x)$ fields are not gauge fields. It can be show that there are $n-1$ mass degenerate $A^{(\underline{k})a}_{\bar n}$ ($\bar n=1,\cdots n-1$) physical scalars, and a massless scalar field $A^{(\underline{k})a}_{\rm G}$, which is recognized as the pseudo-Goldstone boson of $A^{(\underline{k})a}_\mu(x)$, since it can be removed from the theory through a specific NSGT, that is, there is a gauge which maps the gauge $A^{(\underline{k})a}_\mu(x)$ field into a Proca field. \\

To generate the mass spectrum induced by the KK mechanism, we need to specify the set of basis functions  $\{f^{(\underline{0})}_E, \, f^{(\underline{k})}_E(\bar x), f^{(\underline{k})}_O(\bar x)\}$. For this, we need to introduce an observable in the sense of a Hermitian operator that generates an orthogonal vector basis. We use the Casimir invariant $\bar{P}^2=P_{\bar \mu}P_{\bar \mu}$ of the translation group $T(n)$. The generators  $P_{\bar \mu}$ of $T(n)$ induce the orthogonal $\{ \ket {\bar k}\}$ basis through the fundamental equation $P_{\bar \mu}\ket {\bar k}=k_{\bar \mu}\ket {\bar k}$, being these kets also eingenkets of $\bar{P}^2$ with eigenvalues $m^2_{(\underline{k})}\equiv k_{\bar \mu}k_{\bar \mu}$. Note that the ket $\ket {0}$ is associated with an eigenvalue equal to zero. The set of orthogonal functions $\{f^{(\underline{0})}_E, \, f^{(\underline{k})}_E(\bar x), f^{(\underline{k})}_O(\bar x)\}$ emerges after representing the $\{ \ket {\bar k}\}$ basis into the coordinates $\{ \ket {\bar x} \}$ basis, that is, $f^{(\underline{k})}(\bar x)=\braket{\bar x | \bar k}$, with the constant function given by $f^{(\underline{0})}=\braket {\bar x | 0}$.

\subsection{Feynman rules}
In this subsection, we describe the Feynman rules that are needed to calculate the one-loop contributions to both the $H\to \gamma \gamma$ and $H\to gg$ decays in the context of the EDSM. We focus only on those interactions that arise from the compactification of the $(4+n)$-dimensional version of the SM, since, as it is argued in the introduction, they induce the main contributions to physical observables. The bare Lagrangian can be organized as follows:
\begin{eqnarray}
\label{BL}
{\cal L}^{\rm eff}_{\rm B}&=&{\cal L}^{(\underline{0})}_{\rm SM}+{\cal L}^{(\underline{0})}_{\rm GF}+{\cal L}^{(\underline{0})}_{\rm G}
+{\cal L}^{\rm KK}_{(\mathbf{d}\leq 4)}+{\cal L}^{(\underline{k})}_{\rm GF}+{\cal L}^{(\underline{k})}_{\rm G}\nonumber \\
&&+{\cal L}^{\rm KK}_{(\mathbf{d}> 4)}
+{\cal L}^{(\underline{0})}_{\rm c.t.}+{\cal L}^{(\mathbf{d}> 4)}_{\rm c.t.}\, ,
\end{eqnarray}
where the ${\cal L}^{(\underline{0})}_{\rm SM}$ and ${\cal L}^{\rm KK}_{(\mathbf{d}\leq 4)}$ Lagrangians arise from compactification of the $(4+n)$-dimensional version of the SM, so they only have interactions of canonical dimension less than or equal to four, which do not depend on the high energy scale $\Lambda\gg R^{-1}$. In addition, the ${\cal L}^{(\underline{0})}_{\rm GF}+{\cal L}^{(\underline{0})}_{\rm G}$ and ${\cal L}^{(\underline{k})}_{\rm GF}+{\cal L}^{(\underline{k})}_{\rm G}$ Lagrangians correspond to the gauge-fixing and ghost terms associated with the zero-mode gauge fields of the SM and their KK excitations, respectively. The ${\cal L}^{\rm KK}_{(\mathbf{d}> 4)}$ Lagrangian involves interactions of canonical dimension higher than four. Since this Lagrangian emerges from compactification of interactions of canonical dimension higher than $4+n$, such interactions are naturally suppressed by inverse powers of the high energy scale $\Lambda$. Due to the fact that this type of interactions can generate any class of interaction at the tree level, they play a central role in generating the counterterms needed to renormalize amplitudes in a modern sense~\cite{Weinberg}. Here, we focus on those parts of ${\cal L}^{\rm KK}_{(\mathbf{d}> 4)}$ that generate the $H\gamma \gamma$ and $Hgg$ couplings at the tree level. \\

We now proceed to discuss the Feynman rules needed to calculate the contributions induced by the ${\cal L}^{(\underline{0})}_{\rm SM}+{\cal L}^{\rm KK}_{(\mathbf{d}\leq 4)}$ Lagrangians to the $H\gamma \gamma$ and $Hgg$ couplings at the one-loop level.\\

\subsubsection{The fermionic sector}
\ \\
\noindent In the fermion sector, the couplings $\bar{f}^{(\underline{k})}f^{(\underline{k})}A^{(\underline{0})}_\mu$ and $\bar{q}^{(\underline{k})}q^{(\underline{k})}G^{(\underline{0})a}_\mu$ ($f=l,q$) are the same as those of the SM. The coupling $H^{(\underline{0})}\bar{f}^{(\underline{k})}f^{(\underline{k})}$ undergoes a change with respect to its SM analogue. This coupling is given by
\begin{equation}
H^{(\underline{0})}\bar{f}^{(\underline{k})}_{(a)}f^{(\underline{k})}_{(b)}\mapsto -\frac{igm^2_{f^{(\underline{0})}}}{2m_{W^{(\underline{0})}}m_{f^{(\underline{k})}}}
\left[\mathbf{1}-\frac{\Omega}{m_{f^{(\underline{0})}}}\gamma_5\right]_{ab}\, , \ \ \ \ a,b=1,\cdots, 2^{\frac{n}{2}}\, ,
\end{equation}
where $\mathbf{1}$ is the identity matrix of dimension $2^{\frac{n}{2}}\times 2^{\frac{n}{2}}$ and $\Omega$ is a traceless matrix of the same dimension~\cite{EDSM}. However, when this vertex is considered in the fermion-triangle diagrams that leads to the $H\gamma \gamma$ and $Hgg$ couplings, it is found that its contribution is proportional to ${\rm Tr}\{ \mathbf{1}-\frac{\Omega}{m_{f^{(\underline{0})}}}\}={\rm Tr}\{ \mathbf{1}\}=2^{\frac{n}{2}}$.\\

\subsubsection{The bosonic sector}
\ \\
\noindent As far as the boson sector is concerned, the $W^{(\underline{k})-}_\mu W^{(\underline{k})+}_\nu A^{(\underline{0})}_\alpha$ and  $W^{(\underline{k})-}_\mu W^{(\underline{k})+}_\nu A^{(\underline{0})}_\alpha A^{(\underline{0})}_\beta$ vertices coincide with those of the SM. In the case of the $H^{(\underline{0})}W^{(\underline{k})-}_\mu W^{(\underline{k})+}_\nu$ vertex, the corresponding vertex function is identical to that of the SM, that is, it is given by $igm_{W^{(\underline{0})}}g_{\mu \nu}$. However, the presence of a Higgs doublet and its minimal coupling to electroweak gauge fields, together with the occurrence
of the Higgs mechanism at the Fermi scale, renders the mass spectrum of the excited KK scalars $W^{(\underline{k})i}_{\bar \mu}$ and $B^{(\underline{k})}_{\bar \mu}$ rather different from that of a pure Yang-Mills theory. Let   $\phi^{(\underline{0})\dag}=\left(G^{(\underline{0})-}_W,(v+H^{(\underline{0})}-iG^{(\underline{0})}_Z)/\sqrt{2}\right)$ and $\phi^{(\underline{k})\dag}=\left(G^{(\underline{k})-}_W,(H^{(\underline{k})}-iG^{(\underline{k})}_Z)/\sqrt{2}\right)$ be the Higgs doublet and its KK excitations, respectively. Then, once the Higgs mechanism is implemented, some new ingredients emerge, which are not present in pure Yang-Mills  theories. The Higgs mechanism induce mixings between the charged scalars $W^{(\underline{k})\pm}_{\bar \mu}$ and $G^{(\underline{k})\pm}_W$, and, separately, between the neutral scalars $Z^{(\underline{k})}_{\bar \mu}$ and $G^{(\underline{k})}_Z$ as well. However, the scalar KK excitations $G^{(\underline{k})\pm}_W$ and $G^{(\underline{k})}_Z$ mix only with $W^{(\underline{k})\pm}$ and $Z^{(\underline{k})}$, respectively, and they would correspond to massless scalars if no Higgs sector were present. The $n-1$ scalar fields $W^{(\underline{k})\pm}_{\bar n}$ and $Z^{(\underline{k})}_{\bar n}$, with $\bar n=1,\cdots, n-1$ are not mixed with the Higgs doublet components. On the other hand, it can be shown~\cite{EDSM} that the mass matrix that mixes the KK excitations $W^{(\underline{k})\pm}$ and $G^{(\underline{k})\pm}_W$ and the one that mixes $Z^{(\underline{k})}$ and $G^{(\underline{k})}_Z$ have the set of eigenvalues $(m^2_{W^{(\underline{k})}}=m^2_{(\underline{k})}+m^2_{W^{(\underline{0})}},\, 0)$ and $(m^2_{Z^{(\underline{k})}}=m^2_{(\underline{k})}+m^2_{Z^{(\underline{0})}},\, 0)$, respectively. In this way, a physical scalar $W^{(\underline{k})\pm}_n (Z^{(\underline{k})}_n)$ and a massless scalar $W^{(\underline{k})\pm}_{\rm G} (Z^{(\underline{k})}_{\rm G})$ associated with the gauge bosons $W^{(\underline{k})\pm}_\mu$ and $Z^{(\underline{k})}_\mu$ emerge. The massless scalar fields correspond to the pseudo-Goldstone bosons of the vectorial KK excitations of these gauge bosons. So, the $W^{(\underline{0})\pm}_\mu$ and $Z^{(\underline{0})}_\mu$ electroweak gauge bosons have associated a total of $n$ physical scalar fields with masses given by $m_{W^{(\underline{k})}}$  and $m_{Z^{(\underline{k})}}$. This in contrast with the case of gluons or photon, which have associated a total of $n-1$ scalar fields with masses given by $m_{(\underline{k})}$. The couplings of the Higgs boson to the charged scalars, which are needed for our calculations, are given by
\begin{eqnarray}
{\cal L}_{HSS}&=&-gm_{W^{(\underline{0})}}H^{(\underline{0})}\sum_{(\underline{k})}\Bigg[\left(1-\frac{m^2_{H^{(\underline{0})}}}{2m^2_{W^{(\underline{0})}}} \frac{m^2_{(\underline{k})}}{m^2_{W^{(\underline{k})}}}\right)W^{(\underline{k})-}_{ n}W^{(\underline{k})+}_{n}\nonumber \\
&&+\sum^{n-1}_{\bar n=1}W^{(\underline{k})-}_{\bar n}W^{(\underline{k})+}_{\bar n}+
\left(1-\frac{m^2_{H^{(\underline{0})}}}{2m^2_{W^{(\underline{0})}}}\frac{m^2_{W^{(\underline{0})}}}{m^2_{W^{(\underline{k})}}}\right)W^{(\underline{k})-}_{\rm  G}W^{(\underline{k})+}_{\rm G}\Bigg] .
\end{eqnarray}
Notice that, as expected, the Higgs boson distinguishes, from among all scalar fields, the excitations associated with the longitudinal component of the $W$ gauge boson.\\

In this sector, a gauge-fixing procedure for the zero-modes gauge fields of the electroweak sector, $W^{(\underline{0})i}_\mu$ and $B^{(\underline{0})}_\mu$, and their KK excitations,  $W^{(\underline{k})i}_\mu$ and $B^{(\underline{k})}_\mu$, must be introduced. We introduce a nonlinear gauge-fixing procedure which greatly simplifies the loop calculations. We introduce $U_Q(1)$-covariant gauge-fixing functions given by~\cite{Fujikawa,MT,HT,NoTo2,EDYM,OP4}:
\begin{eqnarray}
f^{(\underline{0})\pm}&=&D^{(\underline{0})}_\mu W^{(\underline{0})\pm \mu}-\xi_{(\underline{0})}m_{W^{(\underline{0})}} G^{(\underline{0})\pm}_W, \nonumber \\
f^{(\underline{k})\pm}&=&D^{(\underline{0})}_\mu W^{(\underline{k})\pm \mu}-\xi_{(\underline{k})}m_{W^{(\underline{k})}} W^{(\underline{k})\pm}_{\rm G},
\end{eqnarray}
where $D^{(\underline{0})}_\mu=\partial_\mu -ie A^{(\underline{0})}_\mu$ is the electromagnetic covariant derivative. These gauge-fixing procedure allows us to define the $W^{(\underline{0})\pm}$ and $W^{(\underline{k})\pm}$ propagators in a covariant way under the electromagnetic $U_Q(1,{\cal M}^4)$ gauge group. Since the gauge-fixing Lagrangians ${\cal L}^{(\underline{0})}_{\rm GF}$ and ${\cal L}^{(\underline{k})}_{\rm GF}$ are invariant under the $U_Q(1,{\cal M}^4)$ gauge group, the corresponding ghost Lagrangians will also be invariant under this group. This means that the ghost-antighost contribution will be gauge invariant by itself. Actually, in this gauge, the electromagnetic ghost-antighost couplings resemble scalar electrodynamics. Even more, apart the constant factors that multiply the Higgs-ghost-antighost and the Higgs-scalar-scalar vertices, the one-loop ghost-antighost contribution to $H\gamma \gamma$ is exactly minus twice the corresponding scalar contribution. So, in practice, we only need to calculate the $W$ and scalar (physical or pseudo-Goldstone boson) contribution to obtain the bosonic amplitude associated with the $H\gamma \gamma$ coupling. In both the zero-modes and excited-modes contributions, we will work in the Feynman-'t Hooft gauge ($\xi_{(\underline{0})}=1$, $\xi_{(\underline{k})}=1$).

\section{Decays $H\to \gamma \gamma$ and $H\to gg$}
\label{Ca}
\begin{figure}
\centering\includegraphics[scale=0.8]{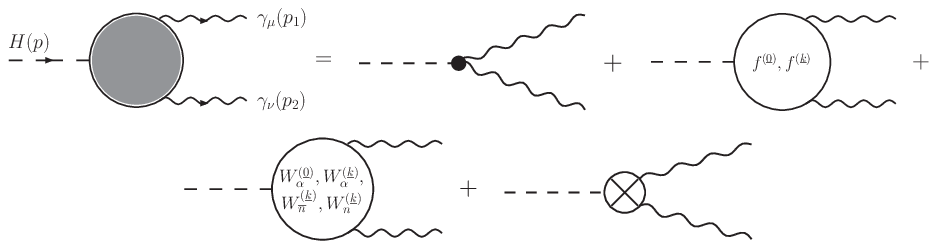}
\caption{\label{Fig1} {\footnotesize}Feynman diagrams that contribute to the $H\gamma \gamma$ coupling in the Feynman-t'Hooft gauge. The tree-level and counterterm diagrams also are displayed.}
\end{figure}

The renormalized invariant amplitude for the $H \to \gamma \gamma$ decay can conveniently be written as follows:
\begin{equation}
\label{HPP}
{\cal M}_{\rm EDSM}(H\to \gamma \gamma)=\frac{i\alpha}{4\pi}\frac{g}{m_{W^{(\underline{0})}}}{\cal A}^{\rm EDSM}_{\gamma \gamma}
\Gamma_{\mu \nu}\epsilon^{\mu *}(p_1,\lambda_1)\epsilon^{\mu *}(p_2,\lambda_2)\, ,
\end{equation}
where
\begin{equation}
\label{APP}
{\cal A}^{\rm EDSM}_{\gamma \gamma}={\cal A}_{\rm tree}^{\gamma \gamma}+{\cal A}^{\gamma \gamma}_{\rm loop}+{\cal A}^{\gamma \gamma}_{\rm c.t.}.
\end{equation}
In the above expressions, $\Gamma_{\mu \nu}=p_{2\mu}p_{1\nu}-\frac{p^2}{2}g_{\mu \nu}$, while ${\cal A}^{\gamma \gamma}_{\rm tree}$, ${\cal A}^{\gamma \gamma}_{\rm loop}$, and ${\cal A}^{\gamma \gamma}_{\rm c.t.}$ represent the tree-level contribution given by interactions of canonical dimension higher than four, the one-loop contribution induced by the ${\cal L}^{(\underline{0})}_{\rm SM}+{\cal L}^{\rm KK}_{(\mathbf{d}\leq 4)}$ Lagrangian, and the counterterm contribution, respectively. This is schematically illustrated in Fig.~\ref{Fig1}. In addition, $p$ is the external momentum associated with the Higgs boson, which we will put at the end on the mass shell, that is, $p^2=m^2_{H^{(\underline{0})}}$. In order to determine the counterterm, we first focus on the one-loop contribution. This one-loop amplitude consists of a fermionic contribution and a bosonic contribution, that is, ${\cal A}^{\gamma \gamma}_{\rm loop}={\cal A}^{\gamma \gamma}_f+{\cal A}^{\gamma \gamma}_b$.\\

\begin{figure}
\centering\includegraphics[scale=0.8]{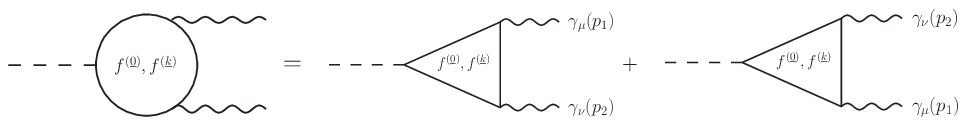}
\caption{\label{Fig2} {\footnotesize Feynman fermionic diagrams that contribute to the $H\gamma \gamma$ coupling in the standard model with universal extra dimensions.}}
\end{figure}

Using the dimensional regularization scheme, the fermionic contribution, which emerges through diagrams shown in Fig.~\ref{Fig2}, is given by:
\begin{eqnarray}
{\cal A}^{\gamma \gamma}_f\Gamma_{\mu \nu}&=&\sum_{f=l,q}\frac{N_cQ^2_f}{m^2_{H^{(\underline{0})}}} \frac{(4\pi \mu^2)^{2-\frac{D}{2}}}{i\pi^{\frac{D}{2}}}\int d^Dk \Bigg\{\frac{T^{(\underline{0})}_{\mu \nu}}{D_{(\underline{0})}}+
2^{\frac{n}{2}}\sum_{(\underline{k})} \frac{T^{(\underline{k})}_{\mu \nu}}{D_{(\underline{k})}}\nonumber \\
&&+\left(\begin{array}{ccc}
\mu \leftrightarrow \nu \\
k_1 \leftrightarrow k_2
\end{array}\right) \Bigg\} ,
\end{eqnarray}
where $\mu$ is the mass scale of the dimensional regularization scheme, $Q_f$ is the electric charge of the fermion $f$ in units of the positron charge, and $N_c$ is the color index, 3 for quarks and 1 for charged leptons. In addition,
\begin{eqnarray}
T^{(\underline{0})}_{\mu \nu}&=&{\rm Tr}\left[(\pFMSlash{k}+m_{f^{(\underline{0})}})(\pFMSlash{k}-\pFMSlash{p}+m_{f^{(\underline{0})}})\gamma^\nu
(\pFMSlash{k}-\pFMSlash{k_1}+m_{f^{(\underline{0})}})\gamma^\mu\right], \\
T^{(\underline{k})}_{\mu \nu}&=&{\rm Tr}\left[(\pFMSlash{k}+m_{f^{(\underline{k})}})(\pFMSlash{k}-\pFMSlash{p}+m_{f^{(\underline{k})}})\gamma^\nu
(\pFMSlash{k}-\pFMSlash{k_1}+m_{f^{(\underline{k})}})\gamma^\mu\right],
\end{eqnarray}
\begin{eqnarray}
D_{(\underline{0})}&=&[k^2-m^2_{f^{(\underline{0})}}][(k-k_1)^2-m^2_{f^{(\underline{0})}}][(k-p)^2-m^2_{f^{(\underline{0})}}], \\
D_{(\underline{k})}&=&[k^2-m^2_{f^{(\underline{k})}}][(k-k_1)^2-m^2_{f^{(\underline{k})}}][(k-p)^2-m^2_{f^{(\underline{k})}}],
\end{eqnarray}
where $m^2_{f^{(\underline{k})}}=m^2_{(\underline{k})}+m^2_{f^{(\underline{0})}}$. Once implemented a Feynman parametrization and integrating over the momenta space, one obtains
\begin{eqnarray}
\label{Af1}
{\cal A}^{\gamma \gamma}_f&=&\sum_f N_c Q^2_f \int^1_0dx \int^{1-x}_0dy \Bigg\{f(x,y)\frac{m^2_{f^{(\underline{0})}}}{\Delta^2_{(\underline{0})F}}\nonumber \\
&&+f(x,y)m^2_{f^{(\underline{0})}}\, 2^{\frac{n}{2}}\sum_{(\underline{k})}\Gamma\left(1+\frac{\epsilon}{2}\right)\left(\frac{1}{4\pi \mu^2}\right)^{-\frac{\epsilon}{2}} \left(\Delta^2_{(\underline{k})F}\right)^{-(1+\frac{\epsilon}{2})}\Bigg\},
\end{eqnarray}
where $f(x,y)=-4\left[1-4(1-x-y)y\right]$, $\Delta^2_{(\underline{0})F}=m^2_{f^{(\underline{0})}}-(1-x-y)yp^2$, $\Delta^2_{(\underline{k})F}=m^2_{(\underline{k})}+\Delta^2_{(\underline{0})F}$, and $\epsilon=4-D$. The symbol $\sum_{(\underline{k})}=\sum_{k_1}\sum_{k_2}\cdots\sum_{k_n}$ represents a multiple sum that runs over every discrete vector $(\underline{k})$ labeling a magnitude $T^{(\underline{k})}$, with the additional restriction that $(\underline{k})\ne(\underline{0})=(0,0,\ldots,0)$. That is, this notation summarizes a total of $ 2^{n}-1 $ different series as follows:
\begin{eqnarray}
\sum_{(\underline{k})}T^{(\underline{k})} & =& \sum_{k_{1}=1}^{\infty}T^{(k_{1},0,\ldots,0)}+\ldots+ \sum_{k_{n}=1}^{\infty}T^{(0,\ldots,k_n)} \nonumber \\
&+&\sum_{k_{1},k_{2}=1}^{\infty}T^{(k_{1},k_{2},0,\ldots,0)}+\ldots+ \sum_{k_{n-1},k_{n}=1}^{\infty}T^{(0,\ldots,0,k_{n-1},k_n)} \nonumber \\
&&\vdots \nonumber \\
 &+&\sum_{k_{1},\ldots, k_{n}=1}^{\infty}T^{(k_{1},\ldots, k_{n})}\ .\label{SD}
\end{eqnarray}
Whereas positions of Fourier indices in the entries of $(\underline{k})$ are not relevant, the number of occupied entries makes a difference. So, in practice, one can use the following definition
\begin{equation}
\label{A2}
\sum_{(\underline{k})}=\sum^n_{l=1}\left(\begin{array}{ccc}
n \\
l
\end{array}\right)\sum^\infty_{k_1=1}\cdots \sum^\infty_{k_l=1}\, .
\end{equation}
Now note that,
\begin{eqnarray}
&&\sum_{(\underline{k})}\left(\frac{1}{4\pi \mu^2}\right)^{-\frac{\epsilon}{2}} \left(\Delta^2_{(\underline{k})F}\right)^{-(1+\frac{\epsilon}{2})}=\frac{1}{R^{-2}}\left(\frac{R^{-2}}{4\pi \mu^2}\right)^{-\frac{\epsilon}{2}}\sum_{(\underline{k})}\left(\underline{k}^2+c^2_F\right)^{-(1+\frac{\epsilon}{2})}\nonumber \\
&=&\frac{1}{R^{-2}}\left(\frac{R^{-2}}{4\pi \mu^2}\right)^{-\frac{\epsilon}{2}}\sum^n_{l=1}\left(\begin{array}{ccc}
n \\
l
\end{array}\right)E^{c^2_F}_l\left(1+\frac{\epsilon}{2}\right),
\end{eqnarray}
where in the last step we have used Eq.~(\ref{A2}), together with the definition of the $l$-dimensional Epstein function~\cite{E1} given by
\begin{equation}
E^{c^2}_l(s)=\sum^\infty_{(k_1,\cdots,k_l)=1}\frac{1}{(k^2_1+\cdots+k^2_l+c^2)^s}.
\end{equation}
In our case, $c^2_F=\frac{\Delta^2_{(\underline{0})F}}{R^{-2}}$. Then, the fermionic contribution (\ref{Af1}) can be written as follows:
\begin{equation}
\label{Af}
{\cal A}^{\gamma \gamma}_f={\cal A}^{(\underline{0})}_f+{\cal A}^{\rm NP}_f,
\end{equation}
where ${\cal A}^{(\underline{0})}_f=\sum_f N_c Q^2_f A^{(\underline{0})}_{\frac{1}{2}}$ is the usual contribution, with the loop function $A^{(\underline{0})}_{\frac{1}{2}}$ given by
\begin{equation}
 A^{(\underline{0})}_{\frac{1}{2}}=-2\tau_{f^{(\underline{0})}}\left[1+(1-\tau_{f^{(\underline{0})}})I^2(\tau_{f^{(\underline{0})}})\right],
\end{equation}
where
\begin{equation}
I(\tau_{f^{(\underline{0})}})=\arctan\left(\frac{1}{\sqrt{\tau_{f^{(\underline{0})}}-1}}\right), \, \, \, \tau_{f^{(\underline{0})}}=\frac{4m^2_{f^{(\underline{0})}}}{p^2}.
\end{equation}
In addition, ${\cal A}^{\rm NP}_f$ represents the new-physics contribution, which is given by:
\begin{eqnarray}
\label{Anpf}
{\cal A}^{\rm NP}_f&=&\sum_f N_c Q^2_f \int^1_0dx \int^{1-x}_0dy f(x,y)\left(\frac{m^2_{f^{(\underline{0})}}}{R^{-2}}\right)\nonumber \\
&&\times \left(\frac{R^{-2}}{4\pi \mu^2}\right)^{-\frac{\epsilon}{2}}\, 2^{\frac{n}{2}}\sum^n_{l=1}\left(\begin{array}{ccc}
n \\
l
\end{array}\right)\Gamma\left(1+\frac{\epsilon}{2}\right)E^{c^2_F}_l\left(1+\frac{\epsilon}{2}\right).
\end{eqnarray}
Note that, in the limit as $\epsilon$ goes to zero, the regularized Epstein function $E^{c^2_F}_l\left(1+\frac{\epsilon}{2}\right)$ diverges for $l=2,3,\cdots$, but it converges for $l=1$.

\begin{figure}
\centering\includegraphics[scale=0.8]{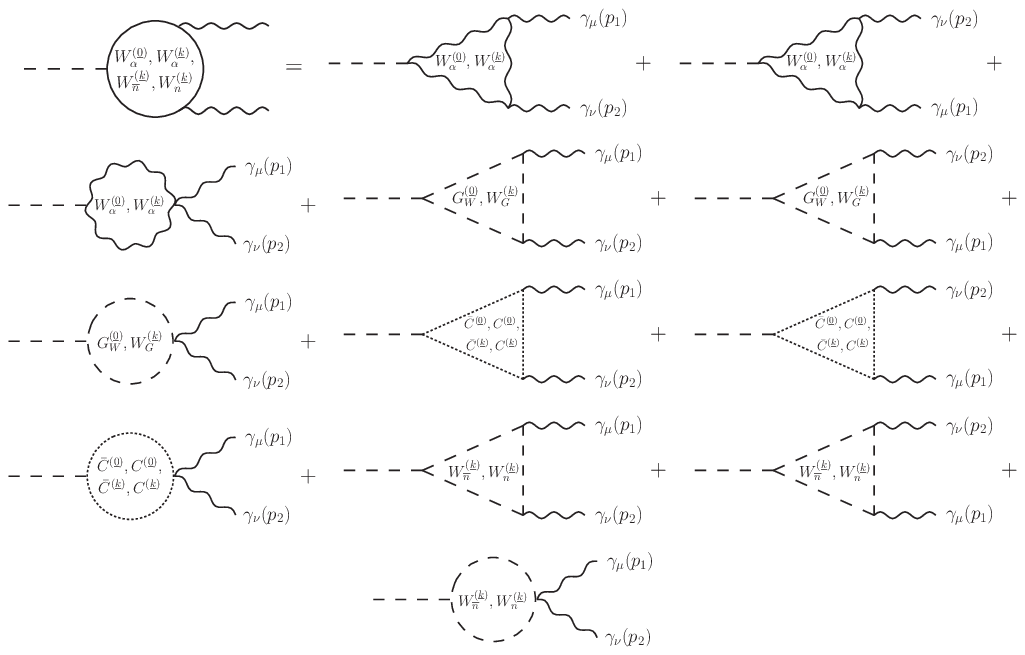}
\caption{\label{Fig3} {\footnotesize Gauge and scalar diagrams that contribute to the $H\gamma \gamma$ coupling in the standard model with universal extra dimensions.}}
\end{figure}

As far as the bosonic contribution is concerned, it is given through diagrams shown in Fig.~\ref{Fig3}. This contribution comprises the gauge contribution and the scalar contribution, that is, ${\cal A}_b={\cal A}_W+{\cal A}_s$. Following the same steps as in the fermionic case, the gauge contribution, which is calculated using the Feynman-'t Hooft version of the nonlinear gauge, can be written as follows:
\begin{eqnarray}
\label{Aw1}
{\cal A}^{\gamma \gamma}_W&=&\int^1_0dx\int^{1-x}_0dy \, g(x,y)\frac{m^2_{W^{(\underline{0})}}}{\Delta^2_{(\underline{0})W}}\Bigg[1 \nonumber \\
&&+\left(\frac{R^{-2}}{4\pi \mu^2}\right)^{-\frac{\epsilon}{2}}c^2_W\sum^n_{l=1}\left(\begin{array}{ccc}
n \\
l
\end{array}\right)\Gamma\left(1+\frac{\epsilon}{2}\right)E^{c^2_W}_l\left(1+\frac{\epsilon}{2}\right)\Bigg],
\end{eqnarray}
where $\Delta^2_{(\underline{0})W}=m^2_{W^{(\underline{0})}}-(1-x-y)yp^2$ and $c^2_W=\frac{\Delta^2_{(\underline{0})W}}{R^{-2}}$. In addition,
\begin{equation}
g(x,y)=8\left[2-\left(3+\frac{m^2_{H^{(\underline{0})}}}{2m^2_{W^{(\underline{0})}}}\right)(1-x-y)y\right].
\end{equation}
Solving the parametric integrals in the first term of (\ref{Aw1}), we have
\begin{eqnarray}
\label{Aw2}
{\cal A}^{\gamma \gamma}_W&=&A^{(\underline{0})}_1+{\cal A}^{\rm NP}_W,
\end{eqnarray}
where $A^{(\underline{0})}_1$ is the usual loop amplitude, which is given by:
\begin{equation}
A^{(\underline{0})}_1=2+3\tau_{W^{(\underline{0})}}+3\tau_{W^{(\underline{0})}}(2-\tau_{W^{(\underline{0})}})I^2(\tau_{W^{(\underline{0})}}),
\end{equation}
and ${\cal A}^{\rm NP}_W$ contains the new-physics effects coming from extra dimensions, given by:
\begin{eqnarray}
\label{Anpw}
{\cal A}^{\rm NP}_W&=&\int^1_0dx\int^{1-x}_0dy \, g(x,y)\left(\frac{m^2_{W^{(\underline{0})}}}{R^{-2}}\right)\nonumber \\
&&\times
\left(\frac{R^{-2}}{4\pi \mu^2}\right)^{-\frac{\epsilon}{2}}\sum^n_{l=1}\left(\begin{array}{ccc}
n \\
l
\end{array}\right)\Gamma\left(1+\frac{\epsilon}{2}\right)E^{c^2_W}_l\left(1+\frac{\epsilon}{2}\right),
\end{eqnarray}
On the other hand, the contribution of the $n$ scalar fields can be written as follows:
\begin{eqnarray}
\label{As}
{\cal A}^{\gamma \gamma}_s&=&\int^1_0dx\int^{1-x}_0dy \, h(x,y) \left(n-\frac{m^2_{H^{(\underline{0})}}}{2m^2_{W^{(\underline{0})}}}\right)\left(\frac{m^2_{W^{(\underline{0})}}}{R^{-2}}\right)\nonumber \\
&&\times \left(\frac{R^{-2}}{4\pi \mu^2}\right)^{-\frac{\epsilon}{2}}\sum^n_{l=1}\left(\begin{array}{ccc}
n \\
l
\end{array}\right)\Gamma\left(1+\frac{\epsilon}{2}\right)E^{c^2_W}_l\left(1+\frac{\epsilon}{2}\right)\,,
\end{eqnarray}
where $h(x,y)=8(1-x-y)y$. Actually, the set of $n$ scalar physical fields $\{W^{(\underline{k})\pm}_{\bar n}, W^{(\underline{k})\pm}_{n}\}$ are genuine KK excitations of the usual $W^{(\underline{0})\pm}_\mu$ gauge field, so it makes sense to jointly consider the contributions of spin-1 and spin-0. Then, we define a bosonic new physics amplitude as follows:
\begin{equation}
{\cal A}^{\rm NP}_b={\cal A}^{\rm NP}_W+{\cal A}^{\gamma \gamma}_s.
\end{equation}

We now turn to present the invariant amplitude for the $H\to gg$ decay. In this case, the one-loop contributions arise only from quarks and their KK excitations. The invariant amplitude can be written as follows:
\begin{eqnarray}
\label{Hgg}
{\cal M}_{\rm EDSM}(H\to gg)&=&\frac{i\alpha_s}{8\pi}\frac{g}{m_{W^{(\underline{0})}}}{\cal A}^{\rm EDSM}_{gg}\Gamma_{\mu \nu}\delta^{ab}\epsilon_a^{\mu *}(p_1,\lambda_1)\epsilon_b^{\mu *}(p_2,\lambda_2)\, ,
\end{eqnarray}
where
\begin{equation}
\label{Agg}
{\cal A}^{\rm EDSM}_{gg}={\cal A}^{gg}_{\rm tree}+{\cal A}^{gg}_{\rm loop}+{\cal A}^{gg}_{\rm c.t.}.
\end{equation}
with ${\cal A}^{gg}_{tree}$, ${\cal A}^{gg}_{loop}$, and ${\cal A}^{gg}_{c.t.}$ representing the tree-level contribution induced by interactions of canonical dimension higher than four, the one-loop contribution induced by interactions of canonical dimension less than or equal to four, and the counterterm contribution, respectively. In this case, the loop amplitude is given by Eq. (\ref{Af}) with $N_c=1$ and $Q^2_f=1$, and the index $f$ in the sum running only over quarks.\\

Assuming that the values of the Higgs boson width do not change appreciably due to the new-physics effects, we can write the observable given by Eq.~(\ref{csw}), for a given number of extra dimensions $n$, as follows:
\begin{eqnarray}
\label{CP}
\mu^{(n)}_{\gamma \gamma}&=&\frac{|{\cal A}^{\rm EDSM}_{gg}|^2}{|{\cal A}^{\rm SM}_{gg}|^2}\frac{|{\cal A}^{\rm EDSM}_{\gamma \gamma }|^2}{|{\cal A}^{\rm SM}_{\gamma \gamma}|^2}\nonumber \, ,\\
&=&P^{(n)}_{gg}C^{(n)}_{\gamma \gamma}\, ,
\end{eqnarray}
where
\begin{eqnarray}
\label{P}
P^{(n)}_{gg}&=&\Bigg | 1+\frac{{\cal A}^{\rm NP}_{gg}}{{\cal A}^{\rm SM}_{gg}}\Bigg |^2 \, ,\\
\label{C}
C^{(n)}_{\gamma \gamma}&=&\Bigg | 1+\frac{{\cal A}^{\rm NP}_{\gamma \gamma}}{{\cal A}^{\rm SM}_{\gamma \gamma}}\Bigg |^2 \, .
\end{eqnarray}
In the above expressions,
\begin{eqnarray}
{\cal A}^{\rm NP}_{gg}&=&{\cal A}^{\rm NP}_{(f=q)}+{\cal A}^{gg}_{\rm tree}+{\cal A}^{gg}_{\rm c.t.}, \\
{\cal A}^{SM}_{gg}&=&\sum_{q}A^{(\underline{0})}_{\frac{1}{2}}\left(\tau_{q^{(\underline{0})}}\right),
\end{eqnarray}
\begin{eqnarray}
{\cal A}^{\rm NP}_{\gamma \gamma}&=&{\cal A}^{\rm NP}_f+{\cal A}^{\rm NP}_b+{\cal A}^{\gamma \gamma}_{\rm tree}+{\cal A}^{\gamma \gamma}_{\rm c.t.},\\
{\cal A}^{\rm SM}_{\gamma \gamma}&=&\sum_{f}A^{(\underline{0})}_{f}\left(\tau_{f^{(\underline{0})}}\right)+{\cal A}^{(\underline{0})}_1.
\end{eqnarray}
In the SM, the top quark dominates the $H\to gg$ decay and gives the most important contribution from the fermion sector to the $H\to \gamma \gamma$ decay, although in the latter process the $W$ contribution is the dominant one. In fact, the top and the $W$ contribute destructively to the  $H\to \gamma \gamma$ decay, the former with an approximate value of $-4/3$ and the latter with $8.35$. So, in absolute value, the top quark contribution represents around the $16\%$ of the $W$ contribution. Using the values $m_{t^{(\underline{0})}}=173$ GeV and $m_{W^{(\underline{0})}}=80.385$ GeV~\cite{PDG}, we have ${\cal A}^{\rm SM}_{gg}\approx -4/3$ and ${\cal A}^{\rm SM}_{\gamma \gamma}\approx 13/2$. Then, Eqs.~(\ref{P}) and (\ref{C}) become
\begin{eqnarray}
P^{(n)}_{gg}&=&\left|1-\frac{4}{3}{\cal A}^{\rm NP}_{gg}\right|^2\, ,\\
C^{(n)}_{\gamma \gamma}&=&\left|1+\frac{2}{13}{\cal A}^{\rm NP}_{\gamma \gamma}\right|^2.
\end{eqnarray}
So the signal strength for a given number $n$ of extra dimensions is given by
\begin{equation}
\mu^{(n)}_{\gamma \gamma}=P^{(n)}_{gg}C^{(n)}_{\gamma \gamma}\, ,
\end{equation}
where $P^{(n)}_{gg}$ would measure the sensitivity to extra dimensions of the Higgs production mechanism, while the $C^{(n)}_{\gamma \gamma}$ parameter would indicate us the corresponding sensitivity of the decay channel.\\

As it is apparent from the results given above, the KK excitations of the top quark and the $W$ gauge boson also dominate the $Hgg$ and $H\gamma \gamma$ couplings, so, from now on, only the contributions from the families of fields $\{t^{(\underline{0})}, t^{(\underline{k})}_{1}, \cdots ,  t^{(\underline{k})}_{2^{\frac{n}{2}}} \}$ and $\{W^{(\underline{0})}_\mu, W^{(\underline{k})}_\mu, W^{(\underline{k})}_{\bar n}, W^{(\underline{k})}_n\}$ will be considered.\\

The contributions induced by tree-level interactions and counterterms, as well as the introduction of an appropriate renormalization scheme, will be discussed below.

\subsection{The $n=1$ case}
As already commented, the case of only one extra dimension deserves special attention. Although the decay $H\to \gamma \gamma$ has already been studied in this context in Ref.~\cite{Petriello}, here, for comparison purposes, we present the main results. We will focus on the $\mu_{\gamma \gamma}$ observable. It is a well-known fact that there is no chirality in odd-dimensional spinor formulations. The construction of the five-dimensional SM requires the symmetry dictated by the orbifold $S^1/Z_2$ used to dimensionally reduce the theory~\cite{OP4,PVSM,OP5}. In five dimensions, as in the four-dimensional case, Dirac fields are still objects with four components. The corresponding generators are given by $S^{MN}=\frac{i}{4}[\gamma^M,\gamma^N]$, with $\gamma^M=\gamma^\mu, i\gamma^5$ the standard Dirac matrices, which satisfy the Clifford's algebra $\{\gamma^M, \gamma^N\}=2g^{MN}$. The generation of the mass terms for fermionic zero modes is somewhat subtle. Its correct implementation leads to a doubly mass-degenerate KK spectrum $f^{(k)}_{(1)}$ and $f^{(k)}_{(2)}$ associated with the zero mode fermion field $f^{(0)}$. In practice, we replace the $2^{\frac{n}{2}}$ factor that multiplies the fermionic loop of the $n\geq2$ case by a factor of $2$.  In the case of only one extra dimension there are no physical scalar fields associated with gluons or the photon. However, there is a physical scalar field associated with each of the $W$ and $Z$ gauge bosons, which arise from a mixing between $W^{(\underline{k})\pm}_5$ and $G^{(\underline{k})\pm}_W$, and a mixing between $Z^{(\underline{k})}_5$ and $G^{(\underline{k})}_Z$, respectively.\\

As emphasized in the Introduction, in this case, the one-loop contribution of KK excitations is proportional to the one-dimensional Epstein function $E^{c^2}_1\left(1+\frac{\epsilon}{2}\right)$, which converges in the $\epsilon \to 0$ limit. So, products of the way $\left(\frac{R^{-2}}{4\pi \mu^2}\right)^{-\frac{\epsilon}{2}}\Gamma\left(1+\frac{\epsilon}{2}\right)E^{c^2}_l\left(1+\frac{\epsilon}{2}\right)$ become $E^{c^2}_1\left(1\right)$. Accordingly, the amplitudes given by Eqs.~(\ref{Af}), (\ref{Aw2}), and (\ref{As}) become
\begin{equation}
\label{Af(1)}
{\cal A}^{(1)}_t={\cal A}^{(\underline{0})}_t+N_cQ^2_t \int^1_0dx \int^{1-x}_0dy f(x,y)\, 2 \, \left(\frac{m_{t^{(\underline{0})}}}{R^{-1}}\right)^2 E^{c^2_F}_1\left(1\right),
\end{equation}
\begin{eqnarray}
\label{Aw2(1)}
{\cal A}^{(1)}_{(W+s)}&=&A^{(\underline{0})}_1+\int^1_0dx\int^{1-x}_0dy \left[g(x,y)+\left(1-\frac{m^2_{H^{(\underline{0})}}}{2m^2_{W^{(\underline{0})}}}\right)h(x,y)\right]\nonumber \\
&&\times \left(\frac{m_{W^{(\underline{0})}}}{R^{-1}}\right)^2 E^{c^2_W}_1\left(1\right),
\end{eqnarray}
where only the contributions from the families of fields $\{t^{(\underline{0})}, t^{(\underline{k})}_{(1)}, t^{(\underline{k})}_{(2)} \}$ and $\{W^{(\underline{0})}_\mu, W^{(\underline{k})}_\mu,  W^{(\underline{k})}_1\}$ were considered. The corresponding amplitude for the $H\to gg$ decay is obtained from expression (\ref{Af(1)}) by putting $N_c=1$ and $Q_t=1$. Since the amplitude is free of ultraviolet divergences, we do not need to consider the contribution of a counterterm. Moreover, the tree level contribution induced by interactions of canonical dimension higher than four is expected to be very suppressed since its strong dependence on the high energy scale $\Lambda$, so we will not consider it.\\

The one-dimensional Epstein function can be expressed in terms of the Riemann zeta function, whose analytical properties are well known in the literature. Such reduction to the Riemann zeta function is given through a power series in $c^2$, which, in our case, means to assume that $c^2=\frac{\Delta^2}{R^{-2}}< 1$, which is valid, since it is expected that the compactification scale $R^{-1}$ to be much larger than the process scale $m_{H^{(\underline{0})}}$. The pass from $E^{c^2}_1(s)$ to $\zeta(s)$ is given by~\cite{EDQED,EDYM,EDST,ER}:
\begin{equation}
\label{EF1}
E^{c^2}_1(s)=\sum^\infty_{k=0}\frac{(-1)^k}{k!}\frac{\Gamma(k+s)}{\Gamma(s)}\zeta(2k+2s)c^{2k}\, ,
\end{equation}
where the Riemman function is defined by
\begin{equation}
\label{RF}
\zeta(s)=\sum^\infty_{n=1}\frac{1}{n^s}\, ,
\end{equation}
which has a simple pole at $s=1$. In our case, Eq.~(\ref{EF1}) becomes
\begin{equation}
E^{c^2}_1(1)=\zeta(2)+F(1,c^2),
\end{equation}
where
\begin{equation}
F(1,c^2)=\sum^\infty_{k=1}\frac{(-1)^k}{k!}\Gamma(k+1)\zeta(2k+2)c^{2k}.
\end{equation}
From these results, one obtains
\begin{eqnarray}
{\cal A}^{\rm NP}_{gg}&=& \int^1_0dx \int^{1-x}_0dy f(x,y)\, 2 \, \left(\frac{m_{t^{(\underline{0})}}}{R^{-1}}\right)^2 \left[\zeta(2)+F(1,c^2_F)\right]\nonumber\\
&=&-\left(\frac{8}{3}\right)\left(\frac{m_{t^{(\underline{0})}}}{R^{-1}}\right)^2 \zeta(2)+\cdots,
\end{eqnarray}
where the ellipsis denotes terms of order $(1/R^{-4})$ and higher. On the other hand, the bosonic amplitude is given by
\begin{eqnarray}
{\cal A}^{\rm NP}_{\gamma \gamma }&=&\int^1_0dx\int^{1-x}_0dy \Bigg\{N_c Q^2_t2f(x,y)\left(\frac{m_{t^{(\underline{0})}}}{R^{-1}}\right)^2\left[\zeta(2)+F(1,c^2_F)\right]+\left(\frac{m_{W^{(\underline{0})}}}{R^{-1}}\right)^2\nonumber \\
&&\times \left[g(x,y)+\left(1-\frac{m^2_{H^{(\underline{0})}}}{2m_{W^{(\underline{0})}}}\right)h(x,y)\right]
\left[\zeta(2)+F(1,c^2_W)\right]
   \Bigg\}\nonumber \\
   &=&\left[-\left(\frac{32}{9}\right)\left(\frac{m_{t^{(\underline{0})}}}{R^{-1}}\right)^2+\frac{1}{3}\left(
   22-\frac{m^2_{H^{(\underline{0})}}}{m^2_{W^{(\underline{0})}}}\right)\left(\frac{m_{W^{(\underline{0})}}}{R^{-1}}\right)^2\right]\zeta(2)+\cdots
\end{eqnarray}

\subsubsection{Discussion}
\ \\
We now proceed to analyze numerically our results for the $n=1$ case. We will fit the values of the compactification scale $R^{-1}$ to a variation range appropriate to reproduce a signal strength $\mu^{(1)}_{\gamma \gamma}$ ranging, approximately, from 1.01 to 1.20. In analyzing our results, we must take into account that the strength signal depends indeed on two subprocesses, namely, $gg\to H$ and $H\to \gamma \gamma$, and that each one may show a different sensitivity to extra dimensions. For instance, it may occur that a very small (large) value of $P^{(1)}_{gg}$ was compensated by a large (small) value of $C^{(1)}_{\gamma \gamma}$, and yet one gets the allowed range of variation $1.01\leq\mu^{(1)}_{\gamma \gamma}\leq1.20$. Our results must also be in accordance with the experimental restriction on the compactification scale, which in the case of one UED is given by $R^{-1}\geq 1.5 \, {\rm TeV}$~\cite{PDG}.\\

\begin{figure}
\centering\includegraphics[scale=1]{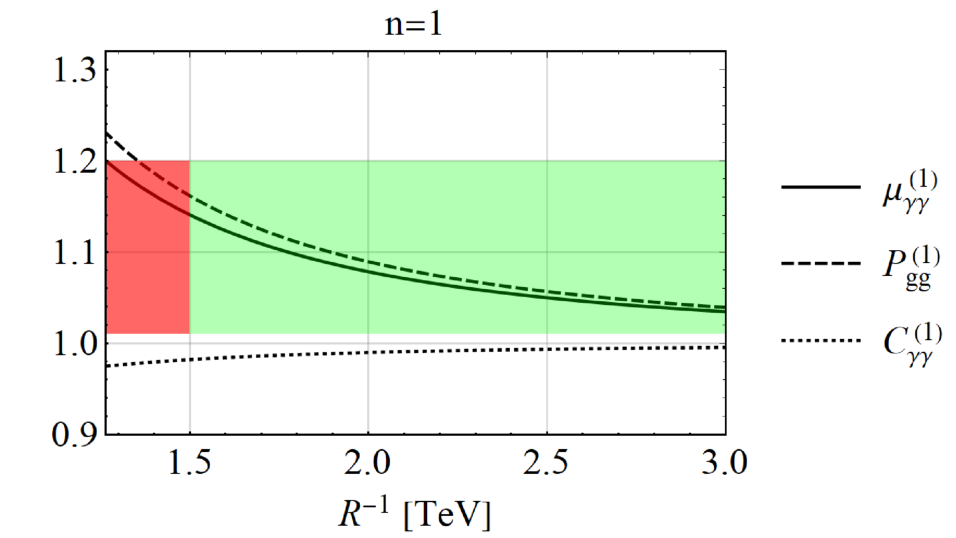}
\caption{\label{MU1} {\footnotesize}The signal strength $\mu^{(1)}_{\gamma \gamma}$ as a function on the compactification scale $R^{-1}$ for the case of only one extra dimension. The relative importance of the subprocesses $ gg \to H$ and $H\to \gamma$ are shown. The allowed region for the experimental constraints $1.01\leq \mu^{(1)}_{\gamma \gamma}\leq 1.20$ and $R^{-1}\geq 1.5\, {\rm TeV}$ are shown.}
\end{figure}

In Fig.~\ref{MU1}, we display the behavior of the quantities $P^{(1)}_{gg}$, $C^{(1)}_{\gamma \gamma}$, and $\mu^{(1)}_{\gamma \gamma}$ as functions on the compactification scale in the range $1.264\,{\rm TeV}\leq R^{-1}\leq 3.0\,{\rm TeV}$. We use as lower value of $R^{-1}$ the one allowed by the experimental restriction on the signal strength $\mu^{(1)}_{\gamma \gamma}$, which is $R^{-1}\geq 1.264 \, {\rm TeV}$. In this figure, the relative importance of the $\{t^{(0)},t^{(k)}_{(1)}, t^{(k)}_{(2)}\}$ quark family and the $\{W^{(0)\pm}_\mu ,W^{(k)\pm}_\mu, W^{(k)\pm} \}$ gauge boson family on the $H\to \gamma \gamma$ decay is also shown. From this figure, it can be appreciated that the new-physics effects interfere constructively with the SM prediction in the production process $gg\to H$ and destructively in the $\gamma \gamma$ channel decay. In the range of variation for $R^{-1}$ shown, $P^{(1)}_{gg}$ ranges from $1.231$ to $1.039$, whereas $C^{(1)}_{\gamma \gamma}$ ranges from $0.975$ to $0.996$, which leads to a variation of the signal strength $\mu^{(1)}_{\gamma \gamma}$ ranging from $1.2$ to $1.034$. \\

From the above results, we can conclude that the signal is determined essentially by the production mechanism, as the channel decay remains practically unchanged due to a strong interference effect between the top and $W$ contributions.\\

\subsection{The $n\geq 2$ case}

In this case, there are Epstein functions of dimension greater than one, which diverge. As it was argued in the introduction, such divergences are associated with short-distance effects in the compact manifold, since they arise from discrete infinite sums. Accordingly, we must define the  interactions of canonical dimension greater than four that generate the counterterms needed for removing such divergences. The corresponding renormalized quantities are contained in the ${\cal L}^{\rm KK}_{(\mathbf{d}>4)}$ and  ${\cal L}^{(\mathbf{d}>4)}_{\rm c.t.}$ Lagrangians. Before defining these interactions, let us analyze more closely the implications of Eq.~(\ref{C2}) for the special case $m=1$. According to Eqs.~(\ref{Anpf}), (\ref{Anpw}), and (\ref{As}), the divergent amplitudes are proportional to
\begin{equation}
\label{GE}
\Phi(c^2,n,\epsilon)\equiv\left(\frac{R^{-2}}{4\pi \mu^2}\right)^{-\frac{\epsilon}{2}}\sum^n_{l=1}\left(\begin{array}{ccc}
n \\
l
\end{array}\right)\Gamma\left(1+\frac{\epsilon}{2}\right)E^{c^2}_l\left(1+\frac{\epsilon}{2}\right)\, .
\end{equation}
Then, using the decomposition of the $l$-dimensional Epstein function in terms of the one-dimensional Epstein function, which is given by Eq.~(\ref{C2}), and expressing the latter one in terms of the Riemann zeta function through Eq.~(\ref{EF1}), the expression (\ref{GE}) becomes
\begin{eqnarray}
\label{GR}
&&\Phi(c^2,n,\epsilon)=\frac{1}{2^{n-1}}\sum^n_{r=1}\sum^r_{l=1}\left(\begin{array}{ccc}
n \\
l-1
\end{array}\right)\pi^{\frac{n-r}{2}}\left(\frac{R^{-2}}{4\pi \mu^2}\right)^{-\frac{\epsilon}{2}}\nonumber \\ &&\times \sum^\infty_{k=0}\frac{(-1)^k}{k!}\Gamma\left(\frac{2k+r-n+2+\epsilon}{2}\right)\zeta(2k+r-n+2+\epsilon)c^{2k}\, .
\end{eqnarray}
The divergences in this series arise as poles of the gamma or Riemann functions. This can happen for an even integer $2k+r-n+2\leq 0$ or for $2k+r-n+2=1$, which correspond to poles, when $\epsilon \to 0$, of the gamma and Riemann functions, respectively. Note that products of the way $\Gamma\left(\frac{2k+r-n+2}{2}\right)\zeta(2k+r-n+2)$ converge for negative even integers $2k+r-n+2=-2,-4,-\cdots$, since they are trivial zeros of the Riemann zeta function. Then, the only pole of the gamma function which leads to a divergence corresponds to $2k+r-n+2=0$. For the analysis that follows, it is convenient to rewrite the above expression so that it explicitly shows those terms that are divergent. Once this rearrangement is implemented, we get
\begin{equation}
\label{Fi1}
\Phi(c^2,n,\epsilon)={\cal A}(c^2,n,\epsilon)+F(n,c^2)\, ,
\end{equation}
where ${\cal A}(c^2,n,\epsilon)$ is the term which contains the divergent parts. It is given by:
\begin{eqnarray}
\label{IF1}
{\cal A}(c^2,n,\epsilon)&=&\sum^{[\frac{n}{2}]}_{k=0}\Bigg\{\left(\frac{R^{-2}}{4\pi \mu^2}\right)^{-\frac{\epsilon}{2}}\Bigg[f_{(k)}(n)\frac{1}{\sqrt{\pi}}
\Gamma\left(\frac{1+\epsilon}{2}\right)\zeta(1+\epsilon)\nonumber \\
 &&+g_{(k)}(n)\Gamma\left(\frac{\epsilon}{2}\right)\zeta(\epsilon)\Bigg]+F_{(k)}(n)  \Bigg\}\nonumber \\
&&\times \sum^k_{j=0}\left(\begin{array}{ccc}
k \\
j
\end{array}\right)\omega^j \left(\frac{m^2}{R^{-2}}\right)^{k-j}\left(\frac{p^2}{R^{-2}}\right)^j,
\end{eqnarray}
where we have used the binomial theorem to express $c^{2k}$ as a polynomial in $\left(\frac{p^2}{R^{-2}}\right)$. In this expression, $m$ stands for $m_{t^{(\underline{0})}}$ or $m_{W^{(\underline{0})}}$, and $\omega=-(1-x-y)y$. Near $\epsilon=0$~\footnote{Note that the pole of the gamma function can be translates to the pole of the zeta function or viceversa using the relation $\Gamma\left(\frac{s}{2}\right)\zeta(s)=\pi^{s-\frac{1}{2}}\Gamma\left(\frac{1-s}{s}\right)\zeta(1-s)$.}, the expression (\ref{IF1}) takes the way
\begin{eqnarray}
\label{IF2}
{\cal A}(c^2,n,\epsilon)&=&\sum^{[\frac{n}{2}]}_{k=0}\Bigg\{
 \frac{1}{2}\left[f_{(k)}(n)-g_{(k)}(n)\right]\left[\frac{2}{\bar{\epsilon}}+\log\left(\frac{\mu^2}{R^{-2}}\right)\right]\nonumber \\
 &&+F_{(k)}(n)\Bigg\}\sum^k_{j=0}\left(\begin{array}{ccc}
k \\
j
\end{array}\right)\omega^j \left(\frac{m^2}{R^{-2}}\right)^{k-j}\left(\frac{p^2}{R^{-2}}\right)^j ,
\end{eqnarray}
where we have used the relations $\Gamma\left(\frac{\epsilon}{2}\right)=\frac{2}{\epsilon}-\gamma +O(\epsilon)$, $\zeta(1+\epsilon)=\frac{1}{\epsilon}+\gamma+O(\epsilon)$, and made some rearrangements so that
\begin{eqnarray}
\label{Eb}
\frac{1}{\bar{\epsilon}}&=&\frac{1}{\epsilon}+\gamma+\frac{1}{2}\log(4\pi)+\frac{1}{2}\psi^{(0)}\left(\frac{1}{2}\right)\nonumber \\
&&+\frac{g_{(k)}(n)}{f_{(k)}(n)-g_{(k)}(n)}\left[\frac{3}{2}\gamma-\log(2\pi)+\frac{1}{2}\psi^{(0)}\left(\frac{1}{2}\right)\right]\, ,
\end{eqnarray}
with $\gamma$ the Euler-Mascheroni constant and $\psi^{(0)}\left(\frac{1}{2}\right)$ the polygamma function of order zero. In addition,
\begin{eqnarray}
F_{(k)}(n)&=&\frac{1}{2^{n-1}}\left(\sum^n_{r\neq n-2k-1,n-2k-2} \right)\sum^r_{l=1}\left(\begin{array}{ccc}
n \\
l-1
\end{array}\right)\pi^{\frac{r-n}{2}}\frac{(-1)^{k}}{k!}\nonumber \\
&&\times \Gamma\left(\frac{2k+r-n+2}{2}\right)\zeta(2k+r-n+2)\, .
\end{eqnarray}
Note that $F_{(k)}(n)$ does  not depend on energy scales, but only on the number $n$ of extra dimensions. On the other hand, $F(n,c^2)$ is a convergent series given by:
\begin{eqnarray}
\label{FFu}
F(n,c^2)&=&\frac{1}{2^{n-1}}\sum^n_{r=1}\sum^{r}_{l=1}\left(\begin{array}{ccc}
n \\
l-1
\end{array}\right)\pi^{\frac{n-r}{2}}\nonumber \\
&&\times \sum^{\infty}_{k=[\frac{n}{2}]+1}\frac{(-1)^{k}}{k!}\Gamma\left(\frac{2k+r-n}{2}\right)\zeta(2k+r-n)c^{2k}\, .
\end{eqnarray}
In the above expressions, the symbol $[\frac{n}{2}]$ means the {\it floor of} $\frac{n}{2}$, that is, the largest integer less than or equal to $\frac{n}{2}$. On the other hand, the functions $f_{(k)}(n)$ and $g_{(k)}(n)$ are given by:
\begin{eqnarray}
\label{fkn}
f_{(k)}(n)&=& \frac{1}{2^{n-1}}\sum^{n-2k-1}_{l=1}\left(\begin{array}{ccc}
n \\
l-1
\end{array}\right)\frac{(-1)^{k}}{k!}\pi^{\frac{2k+1}{2}}\, , \\
\label{gkn}
g_{(k)}(n)&=& \frac{1}{2^{n-1}}\sum^{n-2k-2}_{l=1}\left(\begin{array}{ccc}
n \\
l-1
\end{array}\right)\frac{(-1)^{k}}{k!}\pi^{k+1}\, ,
\end{eqnarray}
which have the following properties:
\begin{eqnarray}
&&f_{(k)}(1)=\cdots=f_{(k)}(2k+1)=0\, , \\
&&g_{(k)}(1)=\cdots=g_{(k)}(2k+2)=0\, .
\end{eqnarray}
These functions can be written in terms of the hypergeometric function $_{2}F_{1}(a,b;c;z)$ as follows:
\begin{eqnarray}
f_{(k)}(n)&=&\frac{2(-1)^k \pi^{\frac{2k+1}{2}}}{\Gamma(k+1)}\Bigg[1-\frac{\Gamma(n+1)}{2^n \Gamma(n-2k)\Gamma(2k+2)}\nonumber \\
&\times&_{2}F_{1}(1,-2k-1;n-2k;-1)\Bigg], \\
g_{(k)}(n)&=&\frac{2(-1)^k \pi^{k+1}}{\Gamma(k+1)}\Bigg[1-\frac{\Gamma(n+1)}{2^n \Gamma(n-2k-1)\Gamma(2k+3)}\nonumber \\
&\times&_{2}F_{1}(1,-2k-2;n-2k-1;-1)\Bigg].
\end{eqnarray}

We now proceed to define interactions of dimension higher than four that render the ${\cal A}^{\rm NP}_{\gamma \gamma}$ and  ${\cal A}^{\rm NP}_{gg}$ amplitudes finite. We propose the following gauge invariant bare Lagrangian:
\begin{eqnarray}
\label{LB1}
{\cal L}^{HGG(WW,BB)}_{\rm B}&=&\sum^{[\frac{n}{2}]}_{k=0}\frac{1}{R^{-2}}\left(\phi^{(\underline{0})\dag}_{\rm B}\left(
\frac{D^2_{\rm B}}{R^{-2}}\right)^k\phi^{(\underline{0})}_{\rm B}\right)\Bigg[\lambda_{{\rm B}(k)}G^{(\underline{0})a}_{{\rm B}\, \mu \nu}G^{(\underline{0})a\mu \nu}_{\rm B} \nonumber \\
&&+\alpha_{{\rm B}(k)}W^{(\underline{0})i}_{{\rm B}\, \mu \nu}W^{(\underline{0})i\mu \nu}_{\rm B} +\beta_{{\rm B}(k)}B^{(\underline{0})}_{{\rm B}\, \mu \nu}B^{(\underline{0})\mu \nu}_{\rm B}\Bigg],
\end{eqnarray}
where $D^2_{\rm B}\equiv D_{{\rm B}\mu}D^\mu_{\rm B}$, with $D_{{\rm B}\mu}$ the bare electroweak covariant derivative, whereas $\lambda_{{\rm B}(k)}$, $\alpha_{{\rm B}(k)}$, and $\beta_{{\rm B}(k)}$ are dimensionless bare constant couplings. The Lagrangian in (\ref{LB1}) arises, after compactification, from its $(4+n)$-dimensional version, so the bare parameters are scaled by powers of $(R^{-1}/\Lambda)$. After moving to renormalized fields and coupling constants through the corresponding renormalization factors, we have
\begin{eqnarray}
\label{LB2}
{\cal L}^{H\gamma \gamma(gg)}_{\rm B}&=&\frac{g}{m_{W^{(\underline{0})}}}\sum^{[\frac{n}{2}]}_{k=0}\left(\frac{v}{R^{-1}}\right)^2\left[\left(
\frac{\partial^2}{R^{-2}}\right)^kH^{(\underline{0})}\right]\Bigg\{\frac{1}{4}\left[\gamma_{(k)}+\delta^{(k)}_\gamma \right]F^{(\underline{0})}_{\mu \nu}
F^{(\underline{0})\mu \nu}\nonumber \\
&&+\frac{1}{2}\left[\lambda_{(k)}+\delta^{(k)}_\lambda\right]\left(\partial_\mu G^{(\underline{0})a}_\nu-\partial_\nu G^{(\underline{0})a}_\mu\right)\partial^\mu G^{(\underline{0})a\nu}
\Bigg\}+\cdots,
\end{eqnarray}
where $\partial^2\equiv \partial_\mu \partial^\mu$, and $\lambda_{(k)}$ and $\gamma_{(k)}$ are renormalized dimensionless parameters, the latter one given by a linear combination of the $\alpha_{(k)}$ and $\beta_{(k)}$ parameters. In addition, $\delta^{(k)}_\lambda$ and $\delta^{(k)}_\gamma$ are the counterterms, and the ellipsis denotes another type of interactions which are not relevant for our study.\\

From the Lagrangian given by Eq.~(\ref{LB2}) and the results shown in Eqs.~(\ref{HPP}), (\ref{APP}), (\ref{Hgg}), and (\ref{Agg}), we can see that the tree-level contributions to the $H\gamma \gamma$ and $Hgg$ interactions are, respectively, given by:
\begin{eqnarray}
{\cal A}^{\gamma \gamma}_{\rm tree}&=&\left(\frac{4\pi}{\alpha}\right)\left(\frac{v}{R^{-1}}\right)^2\sum^{[\frac{n}{2}]}_{k=0}(-1)^k \gamma_{(k)}\left(\frac{p^2}{R^ {-2}}\right)^k\, ,\\
{\cal A}^{gg}_{\rm tree}&=&\left(\frac{4\pi}{\alpha_s}\right)\left(\frac{v}{R^{-1}}\right)^2\sum^{[\frac{n}{2}]}_{k=0}(-1)^k \lambda_{(k)}\left(\frac{p^2}{R^ {-2}}\right)^k\, .
\end{eqnarray}
The renormalized parameters appearing in these amplitudes are quite suppressed, since, as already commented, they are proportional to powers of $\left(R^{-1}/\Lambda\right)$, so, as in the $n=1$ case, we will not consider these types of contributions.\\

On the other hand, to define the counterterms, we need to specify a renormalization scheme. We use a $\overline{\rm MS}$-like scheme, which subtracts, besides the pole at $\epsilon=0$, the numerical constants that usually accompany the $\frac{1}{\epsilon}$ poles. Specifically, we will define counterterms so they remove terms proportional to $\frac{1}{\bar{\epsilon}}$, which is given by Eq.~(\ref{Eb}). So, in the case of the $H\gamma \gamma$ coupling, we define the counterterm contribution as follows:
\begin{eqnarray}
\label{CTPP1}
{\cal A}^{\gamma \gamma}_{\rm c.t.}&=&\frac{4\pi}{\alpha}\left(\frac{v}{R^{-1}}\right)^2\sum^{[\frac{n}{2}]}_{k=0}(-1)^k
\left(\frac{p^2}{R^{-2}}\right)^k\delta^{(k)}_\gamma, \\
\label{CTPP2}
&=&-\frac{\alpha}{4\pi}\int^1_0dx\int^{1-x}_0 dy \sum^{[\frac{n}{2}]}_{k=0}\sum^k_{j=0} \left(\begin{array}{ccc}
k \\
j
\end{array}\right)\omega^j\Bigg\{\nonumber \\
&&2^{\frac{n}{2}}\sum_{f=l,q}N_cQ^2_ff(x,y)\left(\frac{m^2_{f^{(\underline{0})}}}{R^{-2}}\right)
\left(\frac{m^2_{f^{(\underline{0})}}}{R^{-2}}\right)^{k-j}\nonumber \\
&+&\left[g(x,y)+\left(n-\frac{m^2_{H^{(\underline{0})}}}{2m^2_{W^{(\underline{0})}}}\right)\right]
\left(\frac{m^2_{W^{(\underline{0})}}}{R^{-2}}\right)\left(\frac{m^2_{W^{(\underline{0})}}}{R^{-2}}\right)^{k-j}
 \Bigg\}\nonumber \\
 &\times& \left[\frac{1}{2}\left[f_{(k)}(n)-g_{(k)}(n)\right]\frac{2}{\bar{\epsilon}}\right]\left(\frac{p^2}{R^{-2}}\right)^k .
\end{eqnarray}
For a given number $n$ of extra dimensions, the counterterms $\delta^{(k)}_\gamma$ can be determined by comparing coefficients of equal powers of $(p^2/R^{-2})$ in Eqs.~(\ref{CTPP1}) and (\ref{CTPP2}). Similarly, in the case of the $Hgg$ coupling, we define the counterterm contribution as follows:
\begin{eqnarray}
\label{CTgg1}
{\cal A}^{gg}_{\rm c.t.}&=&\frac{8\pi}{\alpha_s}\left(\frac{v}{R^{-1}}\right)^2\sum^{[\frac{n}{2}]}_{k=0}(-1)^k
\left(\frac{p^2}{R^{-2}}\right)^k\delta^{(k)}_\lambda, \\
\label{CTgg2}
&=&-\frac{\alpha_s}{8\pi}\int^1_0dx\int^{1-x}_0 dy \nonumber \\
&\times& \sum^{[\frac{n}{2}]}_{k=0}\sum^k_{j=0} \left(\begin{array}{ccc}
k \\
j
\end{array}\right)\omega^j\Bigg[
2^{\frac{n}{2}}\sum_{q}f(x,y)\left(\frac{m^2_{q^{(\underline{0})}}}{R^{-2}}\right)
\left(\frac{m^2_{q^{(\underline{0})}}}{R^{-2}}\right)^{k-j}\Bigg]\nonumber \\
 &\times& \left[\frac{1}{2}\left[f_{(k)}(n)-g_{(k)}(n)\right]\frac{2}{\bar{\epsilon}}\right]\left(\frac{p^2}{R^{-2}}\right)^k .
\end{eqnarray}

Having removed the divergent parts through amplitudes given by Eqs.~(\ref{CTPP1}-\ref{CTPP2}) and (\ref{CTgg1}-\ref{CTgg2}), the renormalized amplitudes ${\cal A}^{\rm NP}_{\gamma \gamma}$ and ${\cal A}^{\rm NP}_{gg}$ can be written as follows:
\begin{eqnarray}
\label{RAPP1}
{\cal A}^{\rm NP}_{\gamma \gamma}&=&\int^1_0dx \int^{1-x}_0dy \sum^{[\frac{n}{2}]}_{k=0}\sum^k_{j=0} \left(\begin{array}{ccc}
k \\
j
\end{array}\right)\nonumber \\
&\times&\Bigg\{
2^{\frac{n}{2}}\sum_{f=l,q}N_cQ^2_ff(x,y)\left(\frac{m^2_{f^{(\underline{0})}}}{R^{-2}}\right)
\left(\frac{m^2_{f^{(\underline{0})}}}{R^{-2}}\right)^{k-j}\nonumber \\
&+&\left[g(x,y)+\left(n-\frac{m^2_{H^{(\underline{0})}}}{2m^2_{W^{(\underline{0})}}}\right)\right]
\left(\frac{m^2_{W^{(\underline{0})}}}{R^{-2}}\right)\left(\frac{m^2_{W^{(\underline{0})}}}{R^{-2}}\right)^{k-j}
 \Bigg\}\nonumber \\
 &\times& \left\{ \left[\frac{1}{2}\left[f_{(k)}(n)-g_{(k)}(n)\right]\right]\log\left(\frac{\mu^2}{R^ {-2}}\right)+F_{(k)}(n)\right\} \omega^j\left(\frac{p^2}{R^{-2}}\right)^k,
\end{eqnarray}
\begin{eqnarray}
\label{RAgg1}
{\cal A}^{\rm NP}_{gg}&=&\int^1_0dx \int^{1-x}_0dy \sum^{[\frac{n}{2}]}_{k=0}\sum^k_{j=0} \left(\begin{array}{ccc}
k \\
j
\end{array}\right)\nonumber \\
&\times &
2^{\frac{n}{2}}\sum_{q}f(x,y)\left(\frac{m^2_{q^{(\underline{0})}}}{R^{-2}}\right)
\left(\frac{m^2_{q^{(\underline{0})}}}{R^{-2}}\right)^{k-j}
\nonumber \\
 &\times&\left\{ \left[\frac{1}{2}\left[f_{(k)}(n)-g_{(k)}(n)\right]\right]\log\left(\frac{\mu^2}{R^ {-2}}\right)+F_{(k)}(n)\right\}\omega^j\left(\frac{p^2}{R^{-2}}\right)^k.
\end{eqnarray}
The dominant contribution arises from the term $k=0$. Keeping only this contribution, the amplitudes (\ref{RAPP1}) and (\ref{RAgg1}) become
\begin{eqnarray}
\label{RAPP2}
{\cal A}^{\rm NP}_{\gamma \gamma} &=&\int^1_0dx \int^{1-x}_0dy \Bigg\{
2^{\frac{n}{2}}\sum_{f=l,q}N_cQ^2_ff(x,y)\left(\frac{m^2_{f^{(\underline{0})}}}{R^{-2}}\right)
\nonumber \\
&+&\left[g(x,y)+\left(n-\frac{m^2_{H^{(\underline{0})}}}{2m^2_{W^{(\underline{0})}}}\right)\right]
\left(\frac{m^2_{W^{(\underline{0})}}}{R^{-2}}\right)
 \Bigg\}\nonumber \\
 &\times&\left\{ \left[f_{(0)}(n)-g_{(0)}(n)\right]\log\left(\frac{\mu}{R^ {-1}}\right)+F_{(0)}(n)\right\}+\cdots\, ,
\end{eqnarray}
\begin{eqnarray}
\label{RAgg}
{\cal A}^{\rm NP}_{gg}&=&\int^1_0dx \int^{1-x}_0dy \, 2^{\frac{n}{2}}\sum_{q}f(x,y) \left(\frac{m^2_{q^{(\underline{0})}}}{R^{-2}}\right)\nonumber \\
 &\times&
\left\{\left[f_{(0)}(n)-g_{(0)}(n)\right]\log\left(\frac{\mu}{R^ {-1}}\right)+F_{(0)}(n)\right\}+\cdots,
\end{eqnarray}
where the ellipsis denotes terms of order $\left(m^2_{H^{(\underline{0})}}/R^{-2}\right)^2$ and higher.\\

Keeping only the top contribution in the fermion sector and using the definition of the $f_{(k)}(n)$ and $g_{(k)}(n)$ functions given by Eqs.~(\ref{fkn}) and (\ref{gkn}), respectively, we can write:
\begin{eqnarray}
{\cal A}^{\rm NP}_{\gamma \gamma}&=&\left\{-\left(\frac{4}{3}\right)^22^{\frac{n}{2}}
\left(\frac{m^2_{t^{(\underline{0})}}}{m^2_{H^{(\underline{0})}}}\right)+\left[7+\frac{1}{3}
\left(n-\frac{2m^2_{H^{(\underline{0})}}}{m^2_{W^{(\underline{0})}}}\right)
\right]\left(\frac{m^2_{W^{(\underline{0})}}}{m^2_{H^{(\underline{0})}}}\right)\right\}\nonumber \\
&\times&\left(\frac{m^2_{H^{(\underline{0})}}}{R^{-2}}\right)\left\{\left[f_{(0)}(n)-g_{(0)}(n)\right]\log\left(\frac{\mu}{R^{-1}}\right)
+F_{(0)}(n)\right\}\, ,
\end{eqnarray}
\begin{equation}
{\cal A}^{\rm NP}_{gg}=-\left(\frac{4}{3}\right)2^{\frac{n}{2}}
\left(\frac{m^2_{t^{(\underline{0})}}}{R^{-2}}\right)\left\{\left[ f_{(0)}(n)-g_{(0)}(n)\right]\log\left(\frac{\mu}{R^{-1}}\right)+F_{(0)}(n)\right\},
\end{equation}
where
\begin{equation}
 f_{(0)}(n)=2\sqrt{\pi}
\left(1-\frac{n+1}{2^n}\right)\, , \, \, \, g_{(0)}=2\pi\left(1-\frac{n^2+n+2}{2^{n+1}}\right).
\end{equation}

\subsubsection{Discussion}
\ \\
Before discussing our results numerically, the value of the $\mu$ scale must be established. The mass independent renormalization schemes, such as the $\overline{\rm MS}$ scheme which we have introduced here, are not intuitive renormalization schemes, so the $\mu$ scale is, in general, quite arbitrary. However, there are some ways in which $\mu$ can be connected with physical scales of the problem in consideration. Usually, the $\mu$ scale appears as logarithms of the way $\log(\mu/M)$, with $M$ some physical scale of the problem. If we demand our perturbative correction to be small, we must  choose $\mu \sim M$ in order to avoid large logarithms. It is in this sense that the $\mu$ scale can be thought as a physical scale of the problem. In our case, both the ${\cal A}^{\rm NP}_{gg}$ and ${\cal A}^{\rm NP}_{\gamma \gamma}$ amplitudes depend crucially on the quantity
\begin{equation}
\left[f_{(0)}(n)-g_{(0)}(n)\right]\log\left(\frac{\mu}{R^{-1}}\right)+F_{(0)}(n)\, .
\end{equation}
To determine the role played by the term proportional to $\log\left(\frac{\mu}{R^{-1}}\right)$ on physical amplitudes, it is important to know the relative importance of $f_{(0)}(n)-g_{(0)}(n)$ and $F_{(0)}(n)$ as functions of $n$. It turns out that the functions difference $f_{(0)}(n)-g_{(0)}(n)$ is positive for $n\leq 4$ and negative for $n\geq5$, and ranges from $0.8862$ to $-2.4327$ for $n$ varying from 2 to 10. Actually, $f_{(0)}(n)-g_{(0)}(n) \to -2.7382$ for $n\to \infty$, so it is a very slow growing function with $n$. On the other hand, $F_{(0)}(n)$ is a monotonically increasing function of $n$, which takes values in the range $2.5<F_{(0)}(n)<5$ for $n$ varying from $n\geq2$ to arbitrarily large values. Then, if $\mu \sim R^{-1}$ means $\left|\log\left(\frac{\mu}{R^{-1}}\right) \right|<1$, it is clear that the contribution to the physical amplitudes of the $\left[f_{(0)}(n)-g_{(0)}(n)\right]\log\left(\frac{\mu}{R^{-1}}\right)$ term can be ignored compared to the $F_{(0)}(n)$ contribution. In the following, we will work under this assumption.\\

\begin{figure}
\centering\includegraphics[scale=0.5]{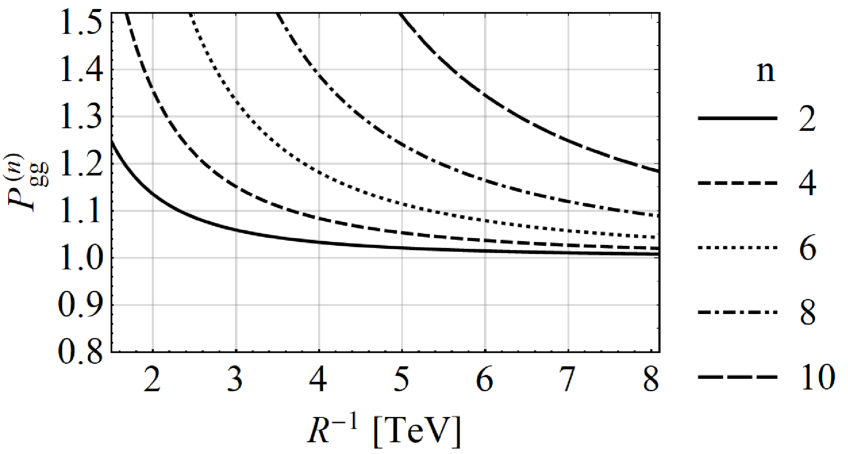}
\centering\includegraphics[scale=0.5]{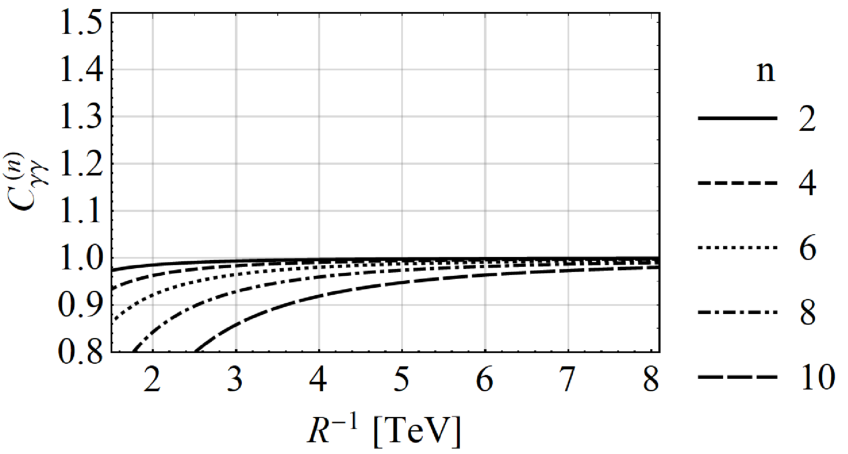}
\centering\includegraphics[scale=1]{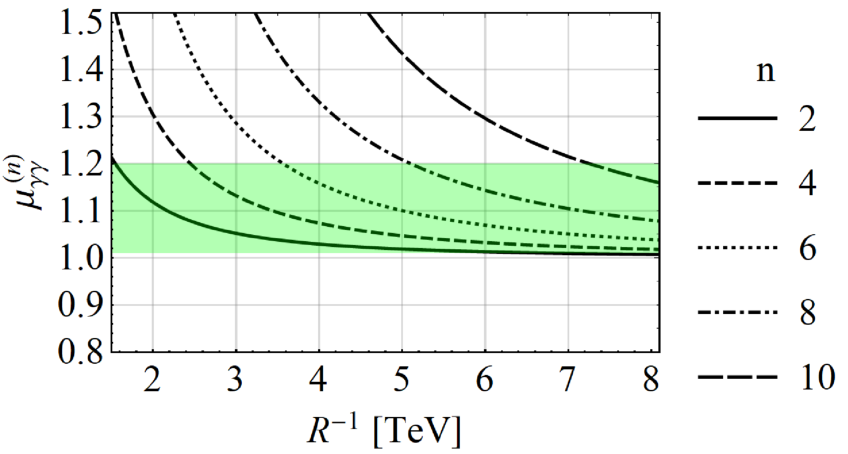}
\caption{\label{Mu} {\footnotesize}The signal strength $\mu^{(n)}_{\gamma \gamma}$ as function on the compactification scale $R^{-1}$ for several values of the number $n$ of extra dimensions. The relative importance of the subprocesses $ gg \to H$ ($P^{(n)}_{gg}$) and $H\to \gamma \gamma$ ($C^{(n)}_{\gamma \gamma}$) are shown. The allowed region for the experimental constraints $1.01\leq \mu^{(n)}_{\gamma \gamma}\leq 1.20$ and $R^{-1}\geq 1.5\, {\rm TeV}$ are shown.}
\end{figure}

In Fig.~\ref{Mu}, the sensitivity to the compactification scale $R^{-1}$ and to the number $n$ of extra dimensions of the subprocesses $ gg \to H$ and $H\to \gamma \gamma$, as well as the signal strength $\mu^{(n)}_{\gamma \gamma}$, is shown. As it can be appreciated from the first graphic in this figure, the subprocess $ gg \to H$ is quite sensitive both to the compactification scale $R^{-1}$ and to the number $n$ of extra dimensions, a fact that is illustrated for the cases $n=2,3,4,6,8,10$. This behavior shows us a constructive effect between the SM prediction and the conribution of extra dimensions, mainly because they have the same sign, which is reflected in a significant impact on the Higgs production through the gluon-fusion mechanism. In contrast, as it can be appreciated from the second graphic in Fig.~\ref{Mu}, things do not  happen the same way in the case of the $H\to \gamma \gamma$ decay channel. The reason for this is that there is a strong  destructive interference between the KK excitations of the top quark and  those of the $W$ gauge boson. As a consequence, the $H\gamma \gamma$ coupling is practically insensitive to effects of extra dimensions. As it can be appreciated from Fig.~\ref{MU1}, this behavior is already observed in the case of only one extra dimension. From second graphic in Fig.~\ref{Mu} and Table~\ref{TABLE}, it can be seen that $C^{(n)}_{\gamma \gamma}<1$ always, and that it takes values very close to the unit for any value of $R^{-1}$ that is compatible with the experimental limit on the signal strength $\mu^{(n)}_{\gamma \gamma}$. In other words, the SM prediction on the $H\to \gamma \gamma$ decay remains essentially unchanged. In contrast, as it can be appreciated from the first graphic in Fig.~\ref{Mu} and from Table~\ref{TABLE}, $P^{(n)}_{gg}>1$ always, and $P^{(n)}_{gg}\to 1$ for $R^{-1} \to \infty$, as it is expected from the decoupling theorem. As it can be seen from the third graphic in Fig.~\ref{Mu}, the net effect leads to scenarios for pairs of quantities $(R^{-1},n)$ that are compatible with the experimental constraint on the signal strength $1.01\leq\mu^{(n)}_{\gamma \gamma}\leq1.2$. This in turn translates into lower limits on the compactification scale $R^{-1}$ that are stronger than the experimental limit $R^{-1}>1.5 \, {\rm TeV}$ derived for the case of only one extra dimension~\cite{PDG}. In Table~\ref{TABLE}, some scenarios allowed by the experimental restriction on $\mu^{(n)}_{\gamma \gamma}$ are shown. We can see from Fig.~\ref{Mu} and Table \ref{TABLE} that the $\mu^{(n)}_{\gamma \gamma}$ observable is highly sensitive to both the compactification scale $R^{-1}$ and the number $n$ of extra dimensions.

\begin{table}[htbp]
\centering
\renewcommand{\arraystretch}{1.5}
\begin{tabular}{|c|c|c|c|c|c|c|}
\hline
$n$  & $R^{-1}\, ({\rm TeV})>$ & $P^{(n)}_{gg}<$ & $C^{(n)}_{\gamma \gamma}>$
\\ \hline
2& $1.548$ & 1.23087 & 0.974903
\\ \hline
$4$ & $2.451$ & 1.23072 & 0.974919
\\ \hline
$6$ & $3.568$ & 1.23080 & 0.974910
\\ \hline
$8$ & $5.102$ & 1.23080 & 0.974910
\\ \hline
$10$ & $7.251$ &1.23084 & 0.974906
\\ \hline
\end{tabular}
\caption{\label{TABLE} Lower bounds on the compactification scale $R^{-1}$ from experimental constraints on the signal strength $\mu^{(n)}_{\gamma \gamma}$ for several values of the number $n$ of extra dimensions. The corresponding highest value of $P^{(n)}_{gg}$ and lowest value of $C^{(n)}_{\gamma \gamma}$ are also shown.}
\end{table}

\section{Conclusions}
\label{Co}Precise measurements of the diverse Higgs-boson decays at future experiments will be crucial in searching for new-physics effects. Special attention deserves the diphoton signal strength, which can be very sensitive to virtual effects of heavy particles, as it is induced at one loop in the SM. In the Higgs resonance, this channel is essentially determined by the decay width $\Gamma(H\to gg)$ and the branching ratio ${\rm BR}(H\to \gamma \gamma)$. In this paper, we have explored the sensitivity of this signal strength to both the size and the dimension of an extra-dimensional compact manifold within the context of the SM with UED.\\

In the context of UED, the one-loop contributions of KK excitations are proportional to discrete and continuous sums, $\sum_{(\underline{k})}\int d^4k$, which can diverge, so they must be regularized. Using the dimensional regularization scheme, it was shown that the regularized discrete sums can naturally be written in terms of multidimensional Epstein functions, with possible divergences emerging from the poles of these functions. While divergences from continuous sums arise through poles of the gamma function and are associated with short-distance effects in the usual spacetime manifold, divergences induced by discrete sums have to do with short-distance effects in the compact manifold. We have argued that divergences arising from discrete sums are genuine ultraviolet divergences in the sense that they correspond to large values of a discrete momentum, which is equivalent to short-distance effects in the compact manifold. In general, we have two types of ultraviolet divergences because we have two different spaces, namely, the usual infinite space and the compact manifold. Therefore, both types of divergences must be treated at the same level in the sense of renormalization theory in a modern sense. In the case at hand, the one-loop amplitudes of the $Hgg$ and $H\gamma \gamma$ couplings are not divergent when only one extra dimension is considered, but they are divergent for $n\geq2$, the divergences arising only from discrete sums, as it is expected from the well-known SM prediction. To define physical amplitudes for these subprocesses, we have introduced interactions of canonical dimension higher than four in order to generate the counterterms needed to renormalize the divergent one-loop amplitudes. We have introduced a $\overline{\rm MS}$-like renormalization scheme that allowed us to remove the divergences appearing through the poles of the gamma and Riemann functions together with some constant quantities that usually accompany such poles. We found that the dominant contribution to renormalized amplitudes is proportional to $\left(\frac{m^2_{H^{(\underline{0})}}}{R^{-2}}\right)\left\{\left[f_{(0)}(n)-g_{(0)}(n)\right]\log\left(\frac{\mu}{R^{-1}}\right)+F_{(0)}(n)\right\}$, but for $\mu \sim R^{-1}$ the $F_{(0)}(n)$ function dominates, so the effect proportional to $\log\left(\frac{\mu}{R^{-1}}\right)$ can be ignored.\\

It is found that the subprocesse $gg\to H$ is quite sensitive to extra dimensions because there is constructive interference between the SM prediction and the extra dimensions effects. However, in the case of the subprocess $H\to \gamma \gamma$, there is a strong destructive interference between the top quark KK excitations and those of the $W$ gauge boson, which leaves practically unchanged the SM prediction. We find that the signal strength $\mu^{(1)}_{\gamma \gamma}$ is quite sensitive to both the size and dimension of the compact manifold, which is a consequence of the high sensitivity of the subprocesse $gg\to H$. We find that in the case of only one extra dimension, the experimental limits on the signal strength $1.01\leq\mu^{(1)}_{\gamma \gamma}\leq1.2$ and on the compactification scale $R^{-1}\geq 1.5 \, {\rm TeV}$  are consistent with each other. However, we find that for $n\geq 2$, consistence with the experimental limit on $\mu^{(n)}_{\gamma \gamma}$ leads to stronger limits on the compactification scale than that reported by the experiment for the case of only one extra dimension.

\section*{Acknowledgments}
We acknowledge financial support from Consejo Nacional de Ciencia y Tecnolog\'\i a (CONACYT) and Sistema Nacional de Investigadores (SNI) (M\' exico). J. M. thanks C\' atedras CONACYT project 1753.\\

\end{document}